\documentclass[prc,preprint,showpacs,preprintnumbers,amsmath,amssymb]{revtex4}

\usepackage{amsmath,amsfonts,amssymb,bm}
\usepackage{graphicx} 
\usepackage{dcolumn}

\usepackage[hang,normalsize,bf]{subfigure} 

\begin{document}


\title{Unintegrated parton distributions \\
and pion production in pp collisions \\ at RHIC's energies} 

\author{M. Czech}
\affiliation{Institute of Nuclear Physics, Polish Academy of Sciences %
PL-31-342 Cracow, Poland} %
\affiliation{Institute of Physics, Jagiellonian University ,%
PL-30-059 Cracow, Poland}
\author{A.Szczurek}
\affiliation{Institute of Nuclear Physics, Polish Academy of Sciences ,%
PL-31-342 Cracow, Poland}
\affiliation{University of Rzesz\'ow, %
PL-35-959 Rzesz\'ow, Poland}

\date{\today}

\begin{abstract}
We compare results of $2 \to 1$ $k_t$-factorization approach
with Kwieci\'nski unintegrated parton distributions
and the standard collinear factorization approach at RHIC and slightly
smaller energies. Our approach contains only one free parameter
responsible for internal parton motion in nucleons.
In contrast to recent works in the literature our $k_t$-factorization
approach includes also quark degrees of freedom in addition
to purely gluonic terms.
Both mid and forward rapidity regions are considered.
We discuss uncertainties due to fragmentation functions.
In general, the $k_t$-factorization approach gives a better description
of the $p_t \sim$ 1 -- 4 GeV region both at mid and forward
rapidity regions.
Our approach leads to asymmetry in the production of $\pi^+$ and
$\pi^-$, very similar to the one observed very recently
by the BRAHMS collaboration.
\end{abstract}


\maketitle

\section{Introduction}

In order to extend the applicability of the collinear-factorization
approach (see e.g. \cite{Owens,Field,EK97,EH02,BS04}) to jet and/or
meson production to small transverse momenta
it was proposed to add an extra Gaussian distribution in the transverse
momentum of colliding partons \cite{CGP78,Owens,Field,BFLPZ01,
Wang2000,Levai,AM04}.
In addition to assuming the parameter of the Gaussian distribution
to be x-independent, one has to substitute
$t \to t - \mu^2$, $u \to u - \mu^2$ and $s \to s + 2 \mu^2$
in order to avoid infinities in the matrix element squared
contributing in this approach at finite transverse momenta.
This procedure is an artifact of the on-shell approximation
(on-shell, non-collinear partons) used.
The result of such a procedure depends on the value of the parameter
$\mu^2$, especially at low transverse momenta.
A part of the arbitrariness can be avoided in the $k_t$-factorization
approach (see e.g. \cite{CS05}).

The $k_t$-factorization approach was used recently to study
both the rapidity and transverse momentum distributions of particles
produced at RHIC energies \cite{KL01,szczurek03} with only
gluon degrees of freedom taken into account.
Recently we have supplemented the mechanisms including the gluon degrees
of freedom with mechanisms including the quark degrees
of freedom \cite{CS05}.
The latter are very important at very forward ($\eta \gg 0$) and
very backward ($\eta \ll 0$) rapidity regions.
Our approach makes use of the recently developed Kwieci\'nski
unintegrated parton distributions \cite{CCFM_b1,CCFM_b2,GKB03}.
In contrast to other approaches in the literature the formalism
takes into account the x-dependent radiative effect
of $k_t$ - broadening of ``initial'' parton distributions.
The formalism of the Kwieci\'nski UPDF is adequate in the region
of not too small longitudinal momentum fractions i.e. at not too
high energies. We have applied these unintegrated parton
distributions to gauge boson \cite{KS04}, standard Higgs \cite{LS05}
hadroproduction as well as for charm-anticharm photoproduction
\cite{LS04}.
Very recently we have applied this formalism also to the description
of the SPS pion production data \cite{CS05}. In the present analysis
we shall apply it to the pion hadroproduction at somewhat larger
energies, up to the RHIC energy W = 200 GeV.

\section{Inclusive cross sections for partons}

The approach proposed by Kwieci\'nski is very convenient to
introduce the nonperturbative effects like
internal (nonperturbative) transverse momentum distributions
of partons in nucleons.
It seems reasonable, at least in the first approximation,
to include the nonperturbative effects in the factorizable way
\begin{equation}
\tilde{f}_i(x,b,\mu^2) = 
\tilde{f}_i^{pert}(x,b,\mu^2)
 \cdot F_i^{np}(b) \; ,
\label{modified_uPDFs}
\end{equation}
where the index $i$ denotes either gluons or quarks or antiquarks.
The form factor responsible for the nonperturbative effects
must be normalized such that \cite{CS05}
\begin{equation}
F_i^{np}(b=0) = 1 \; .
\label{ff_normalization}
\end{equation}
In the following, for simplicity, we use a flavour and
$x$-independent form factor
\begin{equation}
F_g^{np}(b) = F_q^{np}(b) = F_{\bar q}^{np}(b) =  F^{np}(b)
 = \exp\left(-\frac{b^2}{4 b_0^2}\right) \; 
\label{formfactor}
\end{equation}
which describes the nonperturbative effects.
The Gaussian form factor in $b$ means also a Gaussian initial
momentum distribution $\propto \exp(-k_t^2 b_0^2)$ (Fourier transform of
a Gaussian function is a Gaussian function). Gaussian form factor
is often used to correct collinear pQCD calculations for the
so-called internal momenta.
Other functional forms in $b$ are also possible.

In the $k_t$-factorization approach usually the $gg \to g$ fusion
mechanism is included only \cite{KL01}. 
In Ref.\cite{CS05} we have included two other leading-order
diagrams which involve quark degrees of freedom.
They are important in the so-called fragmentation region \cite{CS05}.
The momentum-space formulae for all the processes included read:

for diagram A (gg$\to$g):
\begin{eqnarray}
&&\frac{d \sigma^{A}}{dy d^2 p_t} = \frac{16  N_c}{N_c^2 - 1}
{\frac{1}{p_t^2}}
 \nonumber \\
&& \int
 \alpha_s({\Omega^2}) \;
{  f_{g/1}(x_1,\kappa_1^2,\mu^2)} \;
{  f_{g/2}(x_2,\kappa_2^2,\mu^2)}
\nonumber \\
&&\delta^{(2)}(\vec{\kappa}_1+\vec{\kappa}_2 - \vec{p}_t)
\; d^2 \kappa_1 d^2 \kappa_2    \; ,
\label{diagram_A_tr}
\end{eqnarray}

for diagram B$_1$ (q$_f$ g $\to$ q$_f$):
\begin{eqnarray}
&&\frac{d \sigma^{B_1}}{dy d^2 p_t} = \frac{16  N_c}{N_c^2 - 1}
\left( \frac{4}{9} \right)
{\frac{1}{p_t^2}}
 \nonumber \\
&& \sum_f \int
 \alpha_s({\Omega^2}) \;
{  f_{q_f/1}(x_1,\kappa_1^2,\mu^2)} \;
{  f_{g/2}(x_2,\kappa_2^2,\mu^2)}
\nonumber \\
&&\delta^{(2)}(\vec{\kappa}_1+\vec{\kappa}_2 - \vec{p}_t)
\; d^2 \kappa_1 d^2 \kappa_2    \; ,
\label{diagram_B1_tr}
\end{eqnarray}

for diagram B$_2$ (g q$_f$ $\to$ q$_f$):
\begin{eqnarray}
&&\frac{d \sigma^{B_2}}{dy d^2 p_t} = \frac{16  N_c}{N_c^2 - 1}
\left( \frac{4}{9} \right)
{\frac{1}{p_t^2}}
 \nonumber \\
&& \sum_f \int
 \alpha_s({\Omega^2}) \;
{  f_{g/1}(x_1,\kappa_1^2,\mu^2)} \;
{  f_{q_f/2}(x_2,\kappa_2^2,\mu^2)}
\nonumber \\
&&\delta^{(2)}(\vec{\kappa}_1+\vec{\kappa}_2 - \vec{p}_t)
\; d^2 \kappa_1 d^2 \kappa_2    \; .
\label{diagram_B2_tr}
\end{eqnarray}
These seemingly 4-dimensional integrals can be written
as 2-dimensional integrals after a suitable change
of variables \cite{szczurek03}
\begin{equation}
\int \; ...\; \delta^{(2)}(\vec{\kappa}_1+\vec{\kappa}_2 - \vec{p}_t)
\; d^2 \kappa_1 d^2 \kappa_2 =
\int \; ...\; \frac{d^2 q_t}{4} \; .
\end{equation}
The integrands of these ``reduced'' 2-dimensional integrals in 
$\vec{q}_t = \vec{\kappa_1} - \vec{\kappa_2}$ are
generally smooth functions of $q_t$ and corresponding azimuthal
angle $\phi_{q_t}$.
In Eqs.(\ref{diagram_A_tr}), (\ref{diagram_B1_tr}) and
 (\ref{diagram_B2_tr}) the longitudinal momentum
fractions
\begin{equation}
x_{1/2} = \frac{\sqrt{p_t^2 + m_x^2}}{\sqrt{s}} \exp(\pm y) \; ,
\label{x1_x2}
\end{equation}
where $m_x$ is the effective mass of the parton.
This is important only at $p_t \to 0$ \cite{CS05}.

The sums in (\ref{diagram_B1_tr}) and (\ref{diagram_B2_tr})
run over both quarks and antiquarks.
The argument of the running coupling constant $\Omega^2$ above
was not specified explicitly yet.
In principle, it can be $p_t^2$ or a combination of $p_t^2$,
$\kappa_1^2$ and $\kappa_2^2$. In the standard transverse
momentum representation it is reasonable to assume
$\Omega^2 = \min(p_t^2,\kappa_1^2,\kappa_2^2)$
(see e.g. \cite{szczurek03}). In the region of very small $p_t$ 
usually $p_t^2 < \kappa_1^2, \kappa_2^2$ and $\Omega_2 = p_t^2$
is a good approximation.

Assuming for simplicity that $\Omega^2 = \Omega^2(p_t^2)$ or $p_t^2$
(function of transverse momentum squared of the ``produced'' parton,
or simply transverse momentum squared)
and taking the following representation of the $\delta$ function
\begin{equation}
\delta^{(2)}(\vec{\kappa_1}+\vec{\kappa_2}-\vec{p}_t) =
\frac{1}{(2 \pi)^2} \int d^2 b \;
\exp \left[   
(\vec{\kappa_1}+\vec{\kappa_2}-\vec{p}_t) \vec{b}
\right] \; ,
\label{delta_representation}
\end{equation}
the  formulae (\ref{diagram_A_tr}), (\ref{diagram_B1_tr})
and (\ref{diagram_B2_tr}) can be written in the equivalent way
in terms of parton distributions in the space conjugated
to the transverse momentum.
The corresponding formulae read:

for diagram A:
\begin{eqnarray}
&&\frac{d\sigma^{A}}{dy d^2 p_t} = \frac{16  N_c}{N_c^2 - 1}
{\frac{1}{p_t^2}}
 \alpha_s({p_t^2}) \nonumber \\
&& \int
{\tilde f}_{g/1}(x_1,b,\mu^2) \;
{\tilde f}_{g/2}(x_2,b,\mu^2)
J_0(p_t b) \; 2\pi bdb \; , 
\label{diagram_A_b}
\end{eqnarray}
for diagram B$_1$:
\begin{eqnarray}
&&\frac{d\sigma^{B_1}}{dy d^2 p_t} = \frac{16  N_c}{N_c^2 - 1}
\left( \frac{4}{9} \right)
{\frac{1}{p_t^2}}
 \alpha_s({p_t^2})
 \nonumber \\
&& \sum_f \int
{\tilde f}_{q_f/1}(x_1,b,\mu^2) \;
{\tilde f}_{g/2}(x_2,b,\mu^2)
J_0(p_t b) \; 2\pi bdb \; , 
\label{diagram_B1_b}
\end{eqnarray}
for diagram B$_2$:
\begin{eqnarray}
&&\frac{d \sigma^{B_2}}{dy d^2 p_t} = \frac{16  N_c}{N_c^2 - 1}
\left( \frac{4}{9} \right)
{\frac{1}{p_t^2}}
 \alpha_s({p_t^2})
 \nonumber \\
&& \sum_f \int
{\tilde f}_{g/1}(x_1,b,\mu^2) \;
{\tilde f}_{q_f/2}(x_2,b,\mu^2)
J_0(p_t b) \; 2\pi bdb \; . 
\label{diagram_B2_b}
\end{eqnarray}
These are 1-dimensional integrals. The technical price one has to pay is
that now the integrands are strongly oscillating functions of
the impact factor, especially for large $p_t$.
The formulae (\ref{diagram_A_b}), (\ref{diagram_B1_b})
and (\ref{diagram_B2_b}) are very convenient to directly use
the solutions of the Kwieci\'nski equations discussed
in the previous section.

When extending running $\alpha_s$ to the region of small scales
we use a parameter-free analytic model from ref.\cite{SS97}.

\section{From partons to hadrons}

In Ref.\cite{KL01} it was assumed, based on the concept
of local parton-hadron duality, that the rapidity distribution
of particles is identical to the rapidity distribution of gluons.
In the present approach we follow a different approach
which makes use of phenomenological fragmentation functions (FF's).
In the following we assume $\theta_h = \theta_g$.
This is equivalent to $\eta_h = \eta_g = y_g$, where $\eta_h$ and
$\eta_g$ are hadron and gluon pseudorapitity, respectively. Then
\begin{equation}
y_g = \mathrm{arsinh} \left( \frac{m_{t,h}}{p_{t,h}} \sinh y_h \right)
\; ,
\label{yg_yh}
\end{equation}
where the transverse mass $m_{t,h} = \sqrt{m_h^2 + p_{t,h}^2}$.
In order to introduce phenomenological FF's
one has to define a new kinematical variable.
In accord with $e^+e^-$ and $e p$ collisions we define a 
quantity $z$ by the equation $E_h = z E_g$.
This leads to the relation
\begin{equation}
p_{t,g} = \frac{p_{t,h}}{z} J(m_{t,h},y_h) \; ,
\label{ptg_pth}
\end{equation}
where the jacobian $J(m_{t,h},y_h)$ reads 
\begin{equation}
J(m_{t,h},y_h) =
\left( 1 - \frac{m_h^2}{m_{t,h}^2 \cosh^2 y_h} \right)^{-1/2} \; .
\label{J}
\end{equation}
Now we can write a given-type parton contribution to
the single particle distribution
in terms of a parton (gluon, quark, antiquark) distribution
as follows
\begin{eqnarray}
\frac{d \sigma^{p} (\eta_h, p_{t,h})}{d \eta_h d^2 p_{t,h}} =
\int d y_p d^2 p_{t,p} \int 
dz \; D_{p \rightarrow h}(z,\mu_D^2)  \nonumber \\
\delta(y_p - \eta_h) \; 
\delta^2\left(\vec{p}_{t,h} - \frac{z \vec{p}_{t,p}}{J}\right)
\cdot \frac{d \sigma (y_p, p_{t,p})}{d y_p d^2 p_{t,p}} \; .
\label{from_gluons_to_particles}
\end{eqnarray}

Please note that this is not an invariant cross section.
The invariant cross section can be obtained via suitable
variable transformation
\begin{equation}
\frac{d \sigma^{p} (y_h, p_{t,h})}{d y_h d^2 p_{t,h}} =
\left( \frac{\partial(y_h,p_{t,h})}
            {\partial(\eta_h,p_{t,h})} \right)^{-1} \;
\frac{d \sigma^{p} (y_h, p_{t,h})}{d \eta_h d^2 p_{t,h}} \; ,
\label{invariant_cross_section}
\end{equation}
where
\begin{equation}
y_h = \frac{1}{2} \log
 \left[
\frac{\sqrt{\frac{m_h^2+p_{t,h}^2}{p_{t,h}^2} + \sinh^2\eta_h } + \sinh\eta_h }
     {\sqrt{\frac{m_h^2+p_{t,h}^2}{p_{t,h}^2} + \sinh^2\eta_h } - \sinh\eta_h }
 \right] \; .
\label{yh_etay}
\end{equation}

Making use of the $\delta$ function in
(\ref{from_gluons_to_particles})
the inclusive distributions of hadrons (pions, kaons, etc.)
are obtained through a convolution of inclusive distributions
of partons and flavour-dependent fragmentation functions
\begin{eqnarray}
{
\frac{d \sigma(\eta_h,p_{t,h})}{d \eta_h d^2 p_{t,h}} } =
\int_{z_{min}}^{z_{max}} dz \frac{J^2}{z^2}  \nonumber \\
{D_{g \rightarrow h}(z, \mu_D^2)}
{\frac{d \sigma_{g g \to g}^{A}(y_g,p_{t,g})}{d y_g d^2 p_{t,g}}}
 \Bigg\vert_{y_g = \eta_h \atop p_{t,g} = J p_{t,h}/z}
 \nonumber \\
+ \sum_{f=-3}^3{D_{q_f \rightarrow h}(z, \mu_D^2)}
{\frac{d \sigma_{q_f g \to q_f}^{B_1}(y_{q_f},p_{t,q_f})}
{d y_{q_f} d^2 p_{t,q}}}
 \Bigg\vert_{y_q = \eta_h \atop p_{t,q} = J p_{t,h}/z}
 \nonumber \\
+ \sum_{f=-3}^3{D_{q_f \rightarrow h}(z, \mu_D^2)}
{\frac{d \sigma_{g q_f \to q_f}^{B_2}(y_{q_f},p_{t,q_f})}
{d y_{q_f} d^2 p_{t,q}}}
 \Bigg\vert_{y_q = \eta_h \atop p_{t,q} = J p_{t,h}/z}
 \; . 
\label{all_diagrams}
\end{eqnarray}
One dimensional distributions of hadrons can be obtained
through the integration over the other variable.
For example the pseudorapidity distribution is
\begin{equation}
\frac{d \sigma(\eta_h)}{d \eta_h} =
\int d^2 p_{t,h} \;
\frac{d \sigma(\eta_h,p_{t,h})}{d \eta_h d^2 p_{t,h}} \; .
\label{eta_had_distribution}
\end{equation}
There are a few sets of fragmentation functions available in
the literature (see e.g. \cite{BKK95}, \cite{Kretzer00}, \cite{AKK05}).

\section{Results}

In the present application different fragmentation
functions from the literature \cite{BKK95,Kretzer00,AKK05} will be used.
All of them were obtained in global fits to the $e^+ e^-$ data.
In the present paper we shall show related uncertainties for
pion hadroproduction.

\begin{figure}[htb] 
\begin{center}
\includegraphics[width=6cm]{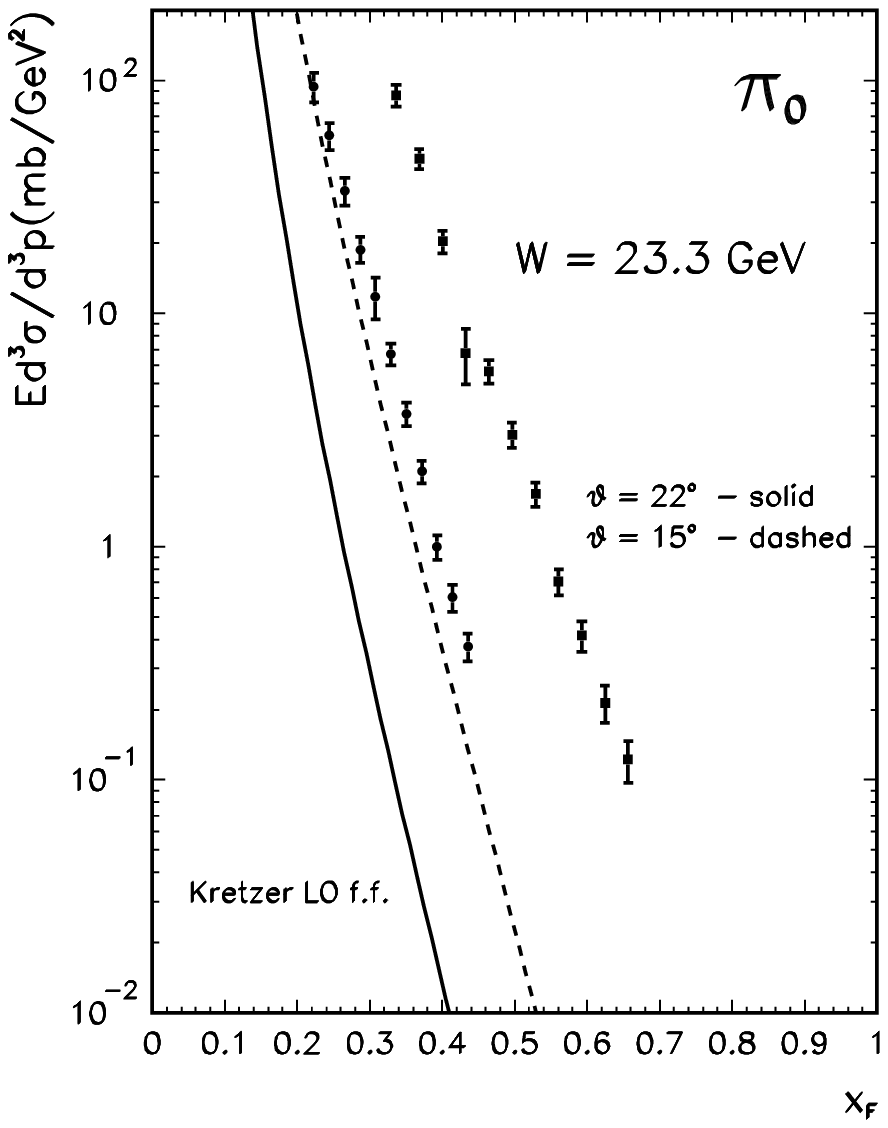}
\includegraphics[width=6cm]{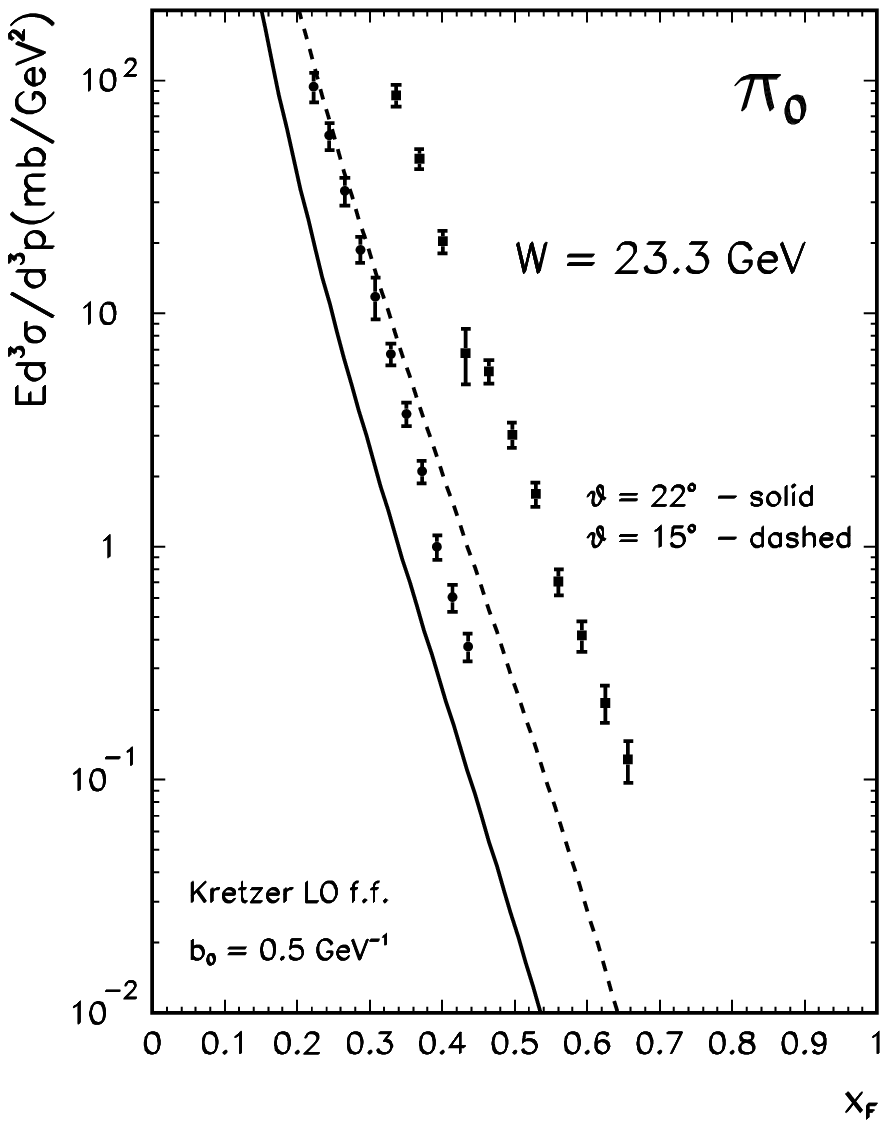}
\caption[*]{
Invariant cross section as a function of $x_F$ at W = 23.3 GeV
for different angles. The standard collinear result is shown in
panel (a) and result of our approach in panel (b).
The experimental data are from \cite{ISR}
\label{fig:ISR23_3}
}
\end{center}
\end{figure}

\begin{figure}[htb] 
\begin{center}
\includegraphics[width=6cm]{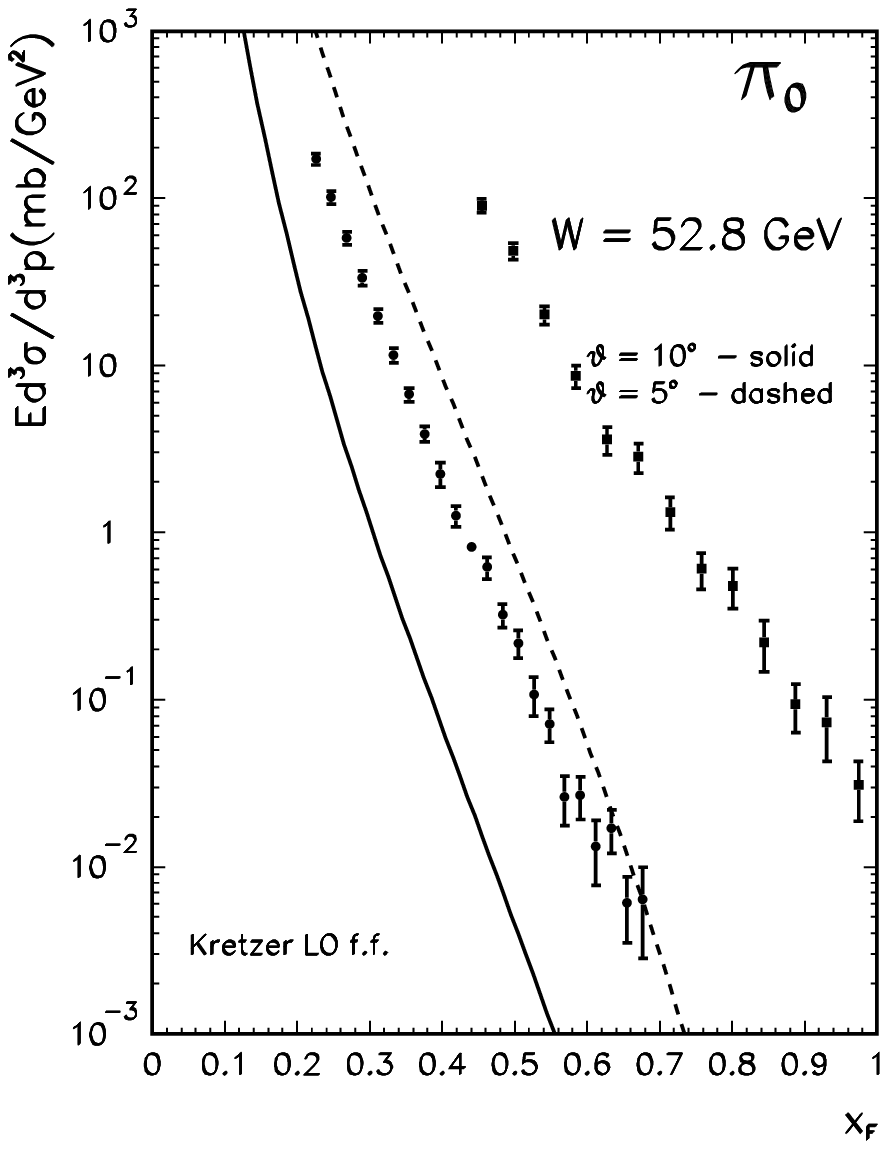}
\includegraphics[width=6cm]{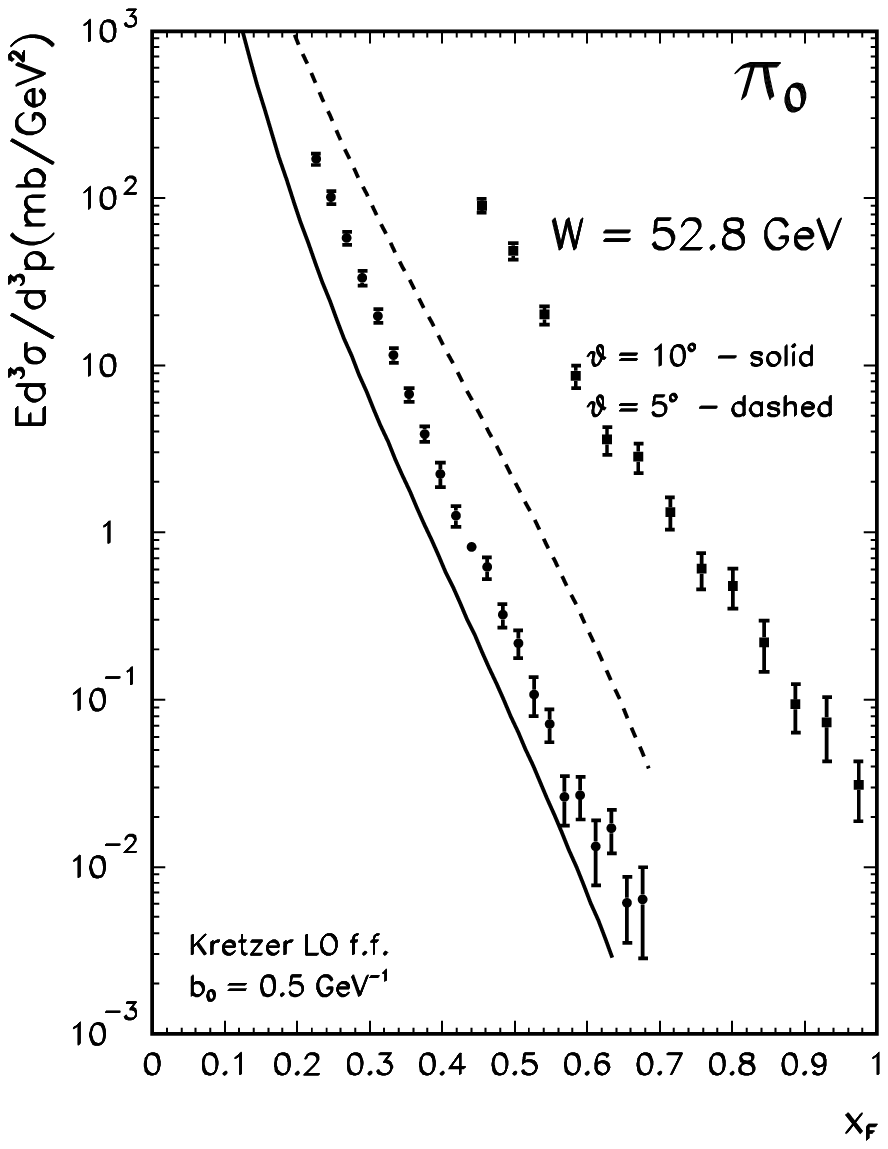}
\caption[*]{
Invariant cross section as a function of $x_F$ at W = 52.8 GeV
for different angles. The standard collinear result is shown in
panel (a) and result of our approach in panel (b).
The experimental data are from \cite{ISR}.
\label{fig:ISR52_8}
}
\end{center}
\end{figure}

Before we go to the RHIC data we wish to look at some lower-energy
data measured at ISR \cite{ISR}.
It was pointed out recently \cite{BS04} that the standard
collinear approach is not able to describe very forward
production of $\pi^0$.
In Fig.\ref{fig:ISR23_3} we present invariant cross section
as a function of the Feynman $x_F$ for W = 23.3 GeV at
two different laboratory angles $\theta$ = 15$^o$, 22$^o$.
The collinear factorization result (left panel) lies well below the
data independent of what fragmentation functions are used.
The results of the $k_t$-factorization approach describe
the experimental data much better.
In Fig.\ref{fig:ISR52_8} we present similar results for somewhat
larger energy W = 52.8 GeV and $\theta$ = 5$^o$, 10$^o$.
The situation here is very similar to that at the lower energy.

Clearly, the standard collinear approach fails badly in the very
forward region. The $k_t$-factorization is much better but still
some contribution is missing at large Feynman $x_F$ and at 
small transverse momenta. However, one should remember that other
processes such as pion stripping \cite{pion_stripping} and/or diffractive
production of nucleon resonances and their subsequent decay may play
an important role here. This requires a separate analysis
which goes beyond the scope of the present paper.

Let us return to the midrapidity region.
The PHENIX collaboration has measured invariant cross section
as a function of the $\pi^0$ transverse momentum at W = 200 GeV
in a very narrow interval of pseudorapidity $\eta$ = 0.0 $\pm$ 0.15.

In Fig.\ref{fig:PHENIX_FF} we show our full result
(diagrams $A$, $B_1$ and $B_2$ \cite{CS05})
for different fragmentation functions \cite{BKK95,Kretzer00,AKK05}.
In this calculation $b_0$ = 0.5 GeV$^{-1}$ was used.
This is the optimal value of the parameter for gauge boson production
\cite{KS04}.
\begin{figure}[htb] 
\begin{center}
\includegraphics[width=7cm]{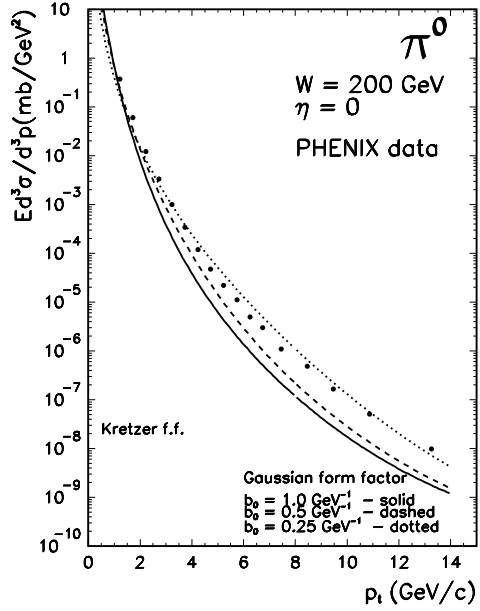}
\caption[*]{
Invariant cross section for $\pi^0$ production as a function
of pion transverse momentum at W = 200 GeV and $\eta$ = 0.0.
The $k_t$-factorization results are shown for different
values of the parameter $b_0$.
The experimental data of the PHENIX collaboration are
from \cite{PHENIX_pi0}.
\label{fig:PHENIX_b0}
}
\end{center}
\end{figure}
\begin{figure}[htb] 
\begin{center}
\includegraphics[width=7cm]{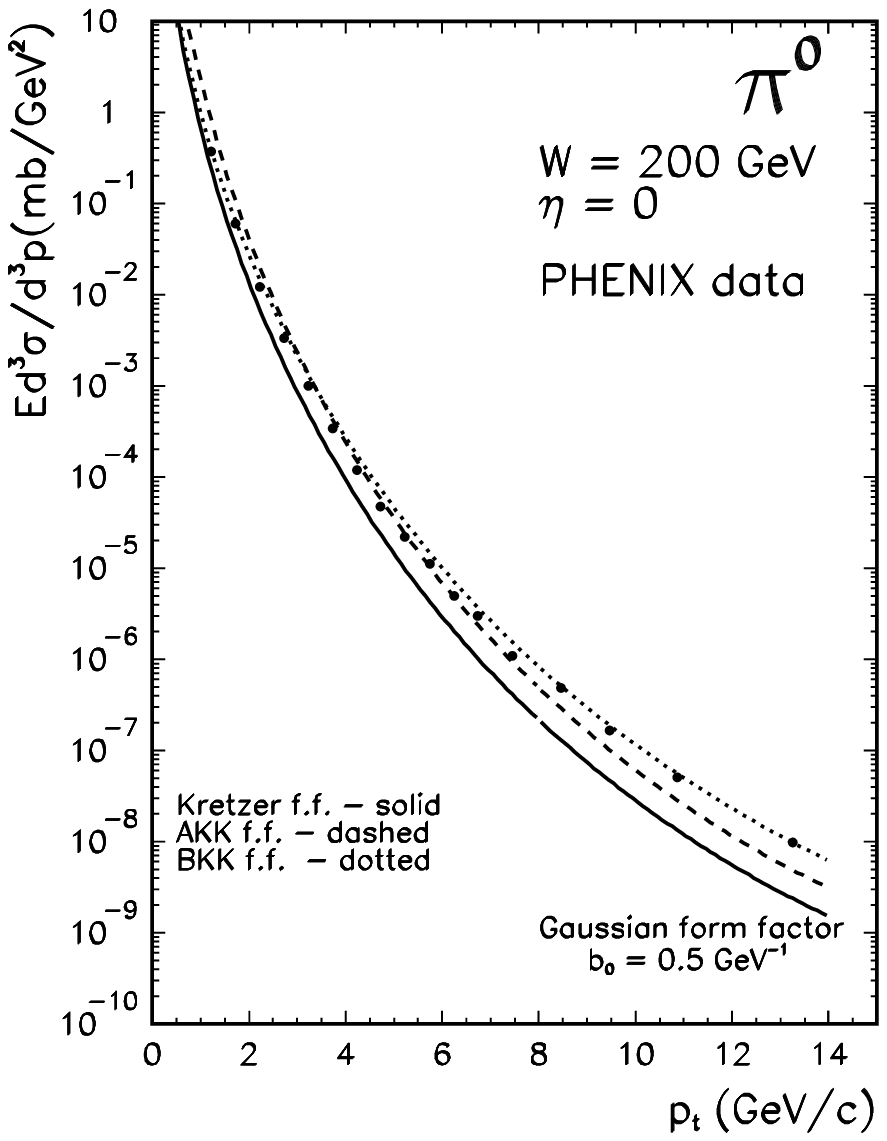}
\caption[*]{
Invariant cross section for $\pi^0$ production as a function
of pion transverse momentum at W = 200 GeV and $\eta$ = 0.0.
The $k_t$-factorization results are shown
for different sets of fragmentation functions.
The experimental data of the PHENIX collaboration are
from \cite{PHENIX_pi0}.
\label{fig:PHENIX_FF}
}
\end{center}
\end{figure}
In Fig.\ref{fig:PHENIX_b0} we show the dependence on the value of
the parameter $b_0$.
Having in view the uncertainties in the fragmentation functions and
in the parameter $b_0$, which is responsible for nonperturbative
effects like e.g. parton Fermi motion, 
our $k_t$-factorization result describes the data very well and
in a quite broad range of transverse momentum.
In Fig.\ref{fig:PHENIX_contributions} we show individual contributions
of gluon and quark components. In contrast to the standard beliefs
(see e.g. \cite{KL01}) the quark contributions are only slightly
smaller then the gluon ones and certainly not negligible,
especially at larger transverse momenta.
\begin{figure}[htb] 
\begin{center}
\includegraphics[width=5cm]{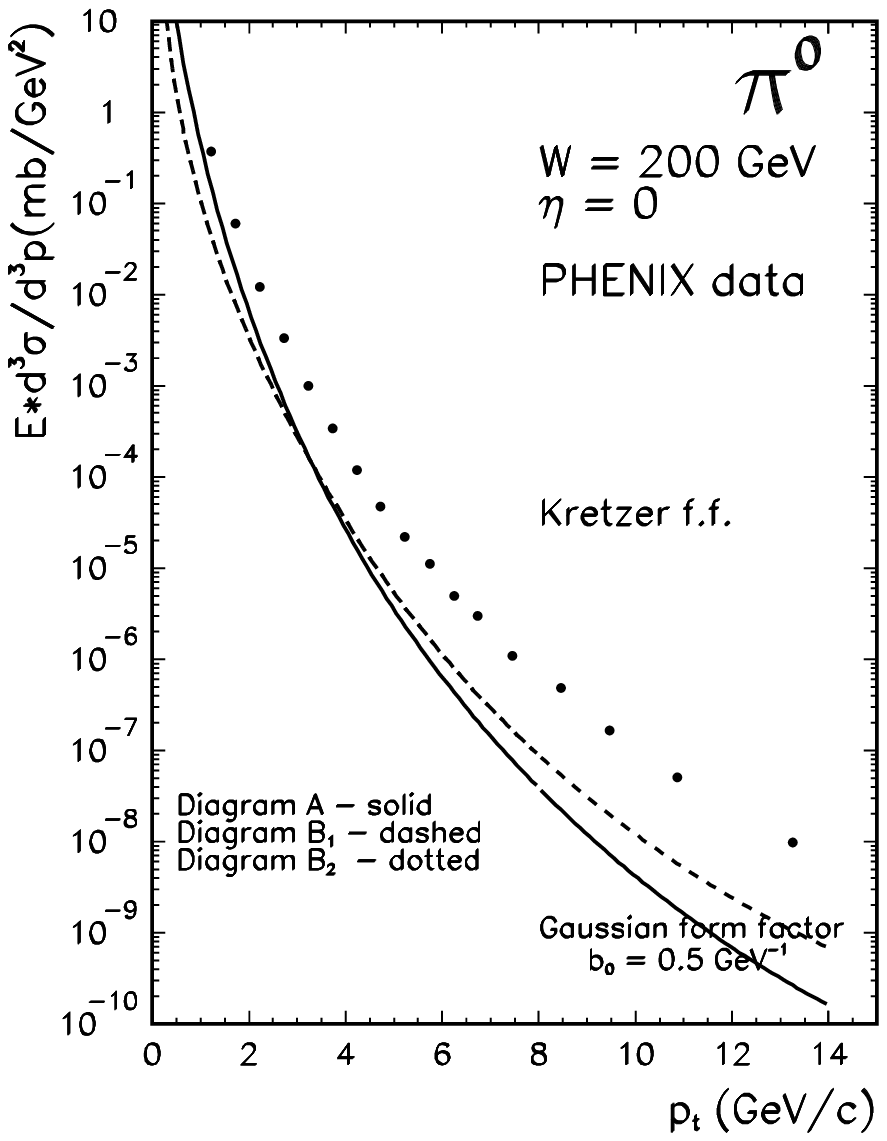}
\includegraphics[width=5cm]{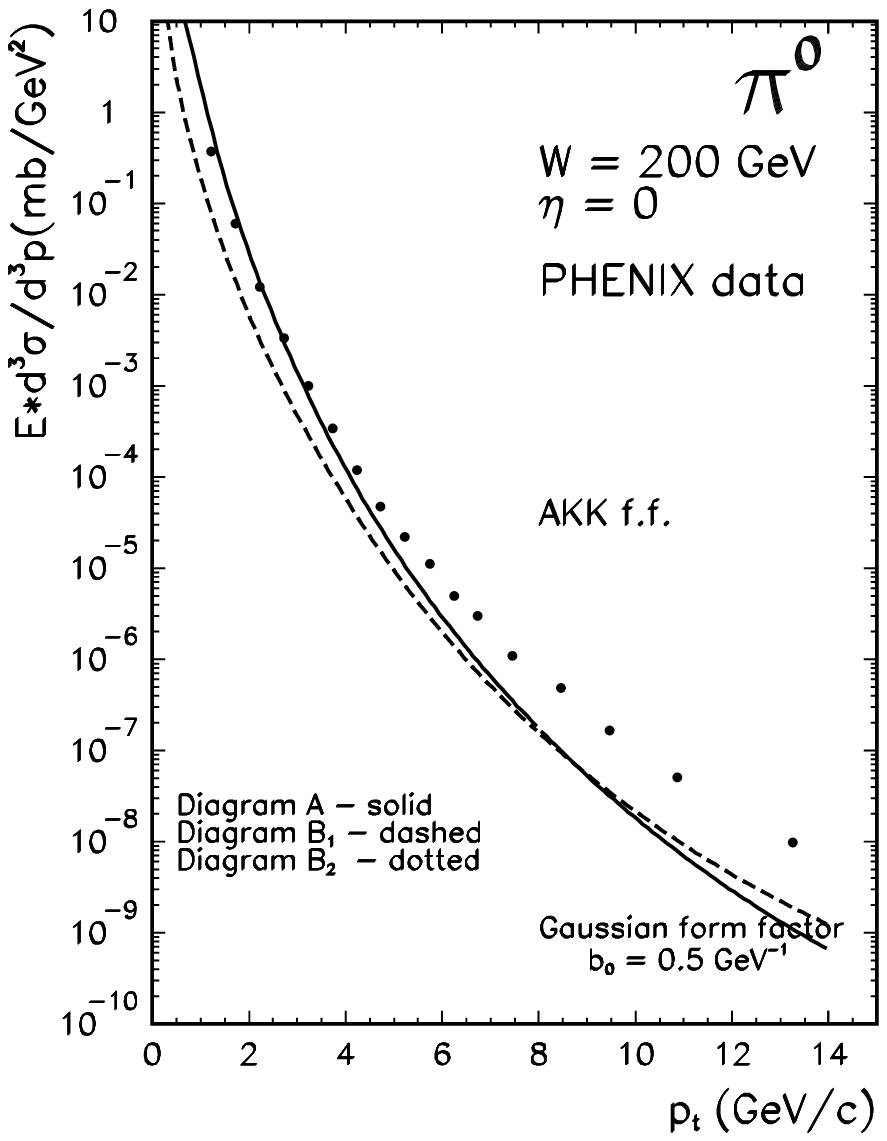}
\includegraphics[width=5cm]{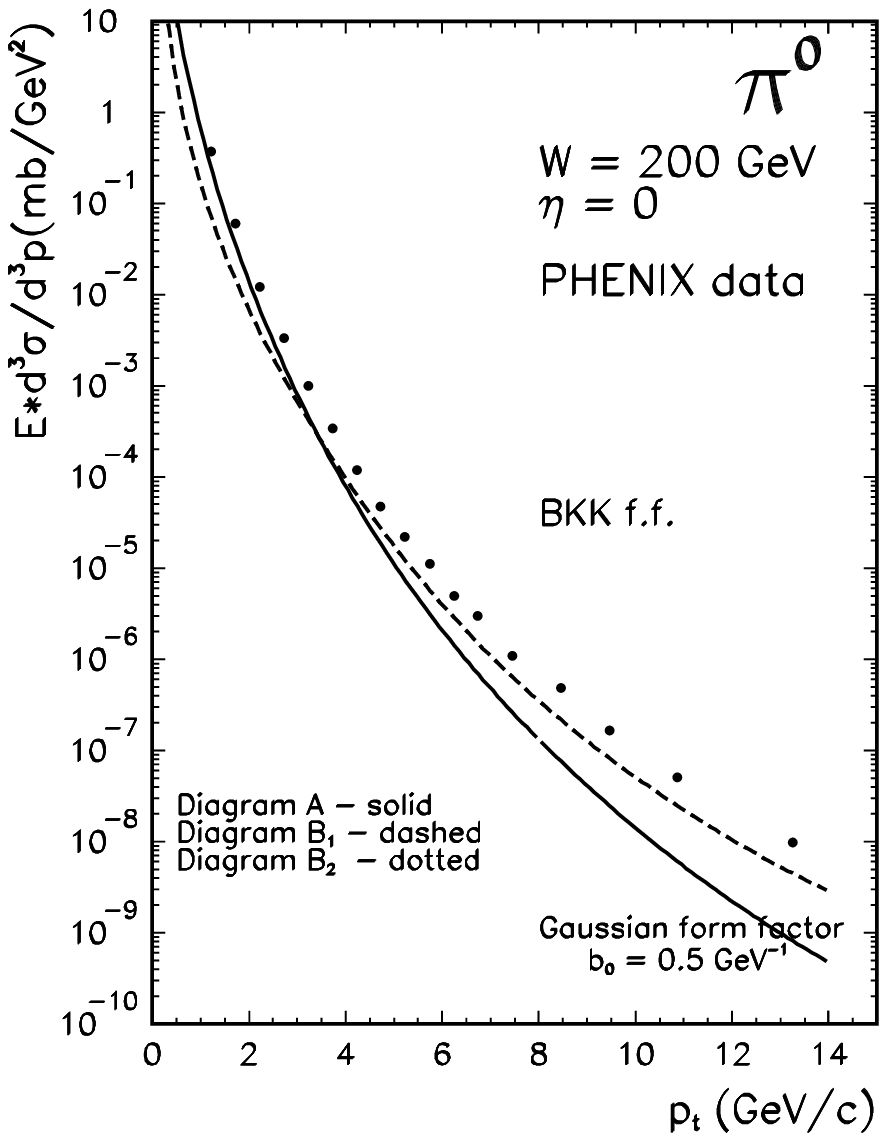}
\caption[*]{
Invariant cross section for $\pi^0$ production as a function
of pion transverse momentum at W = 200 GeV and $\eta$ = 0.0.
Individual contributions of gluon and quark
fragmentation are shown for different fragmentation functions.
The experimental data of the PHENIX collaboration are
from \cite{PHENIX_pi0}.
\label{fig:PHENIX_contributions}
}
\end{center}
\end{figure}

\begin{figure}[htb] 
\begin{center}
\includegraphics[width=4cm]{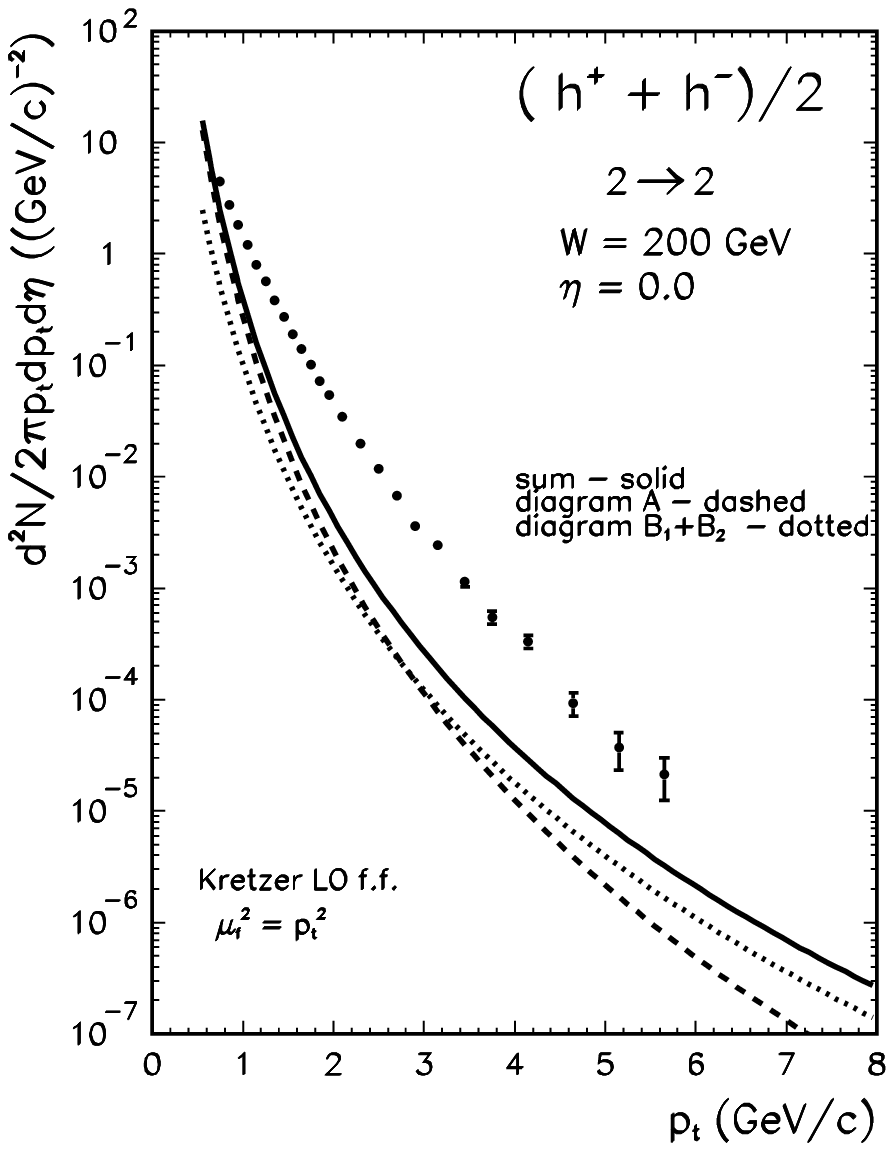}
\includegraphics[width=4cm]{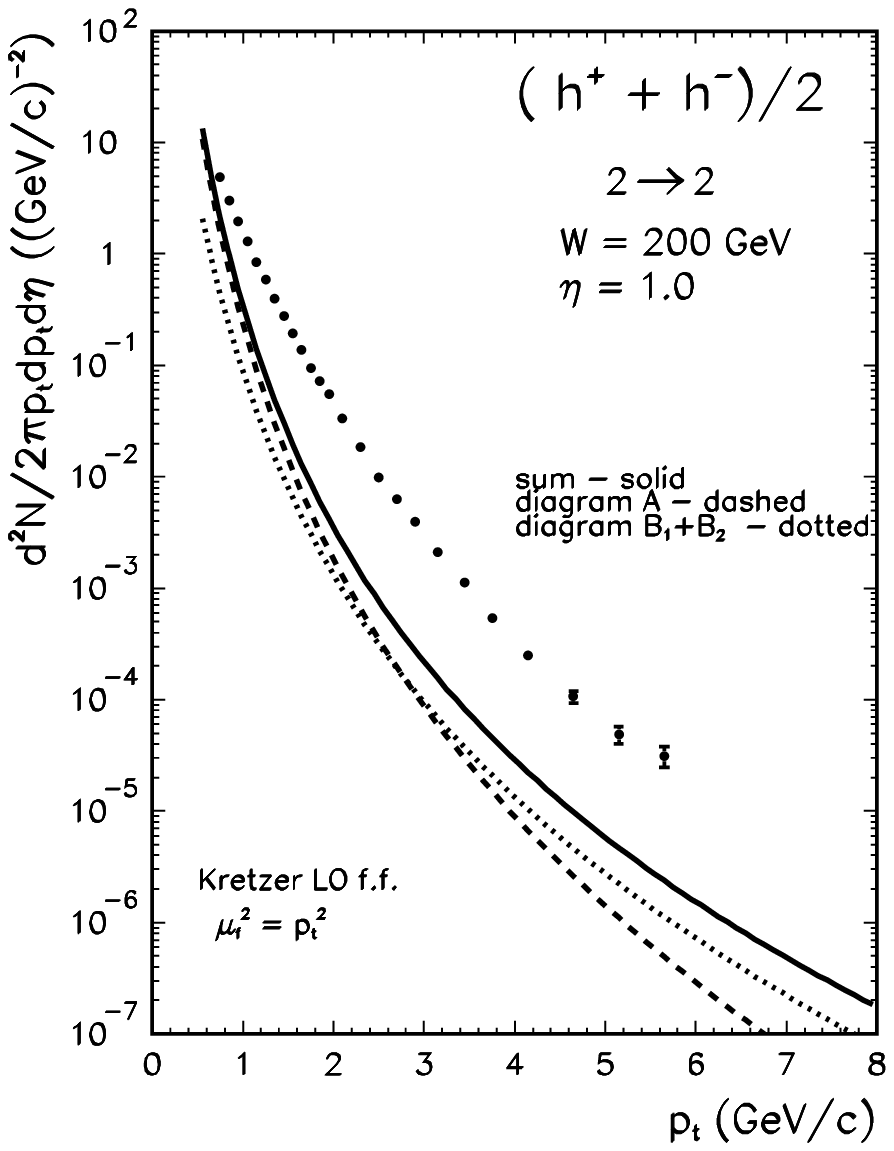}
\includegraphics[width=4cm]{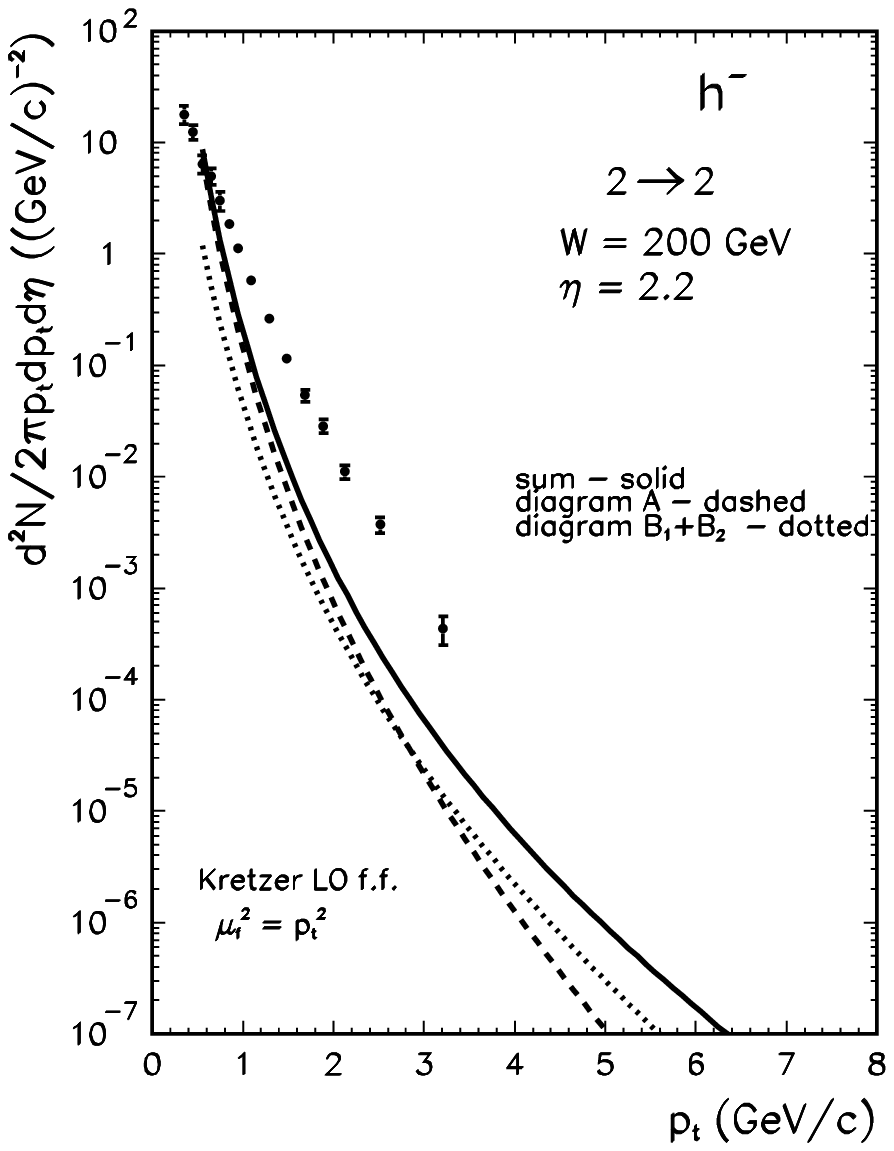}
\includegraphics[width=4cm]{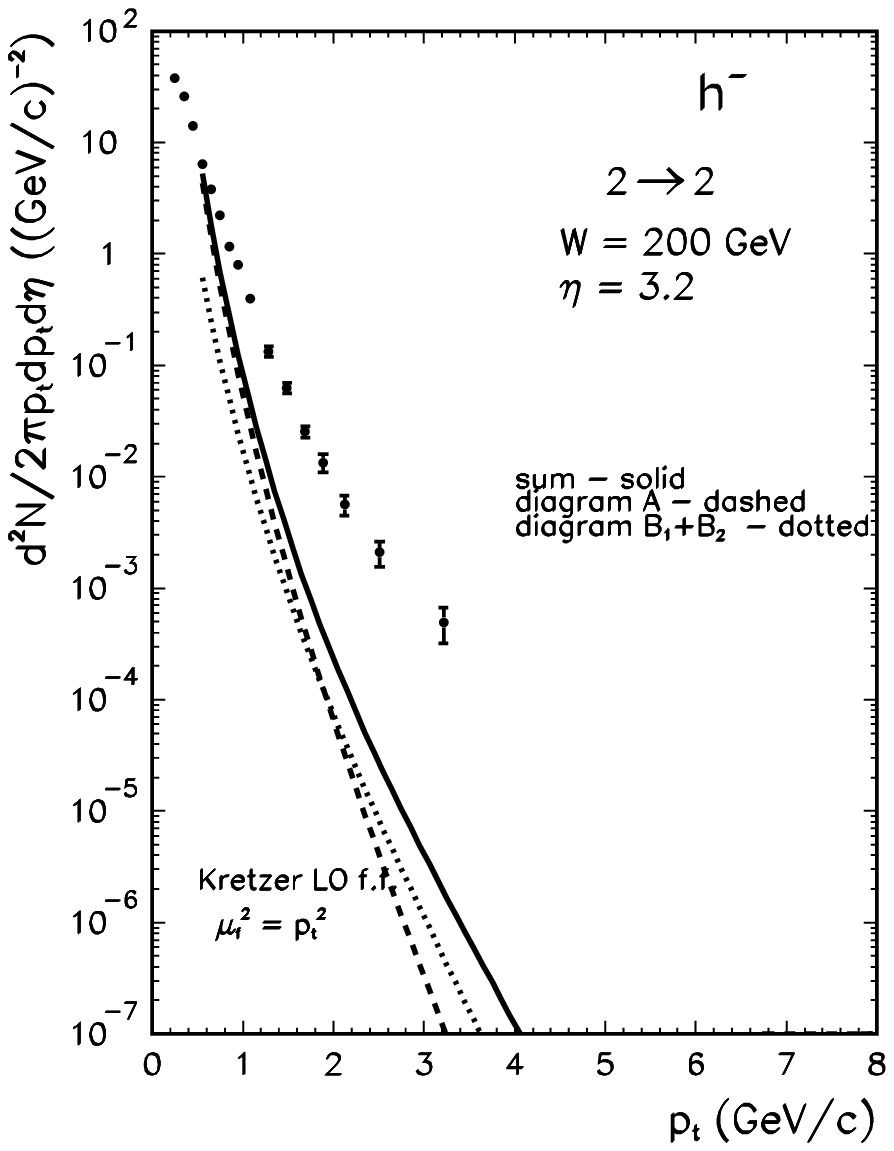}
\caption[*]{
Invariant cross section for charged particle production for
different values of particle pseudorapidity at W = 200 GeV.
The lines represent results obtained within standard collinear
factorization approach and Kretzer fragmentation functions
\cite{Kretzer00}.
The BRAHMS collaboration experimental data
\cite{BRAHMS_charged_hadrons}
are shown by the solid circles.
\label{fig:BRAHMS_coll}
}
\end{center}
\end{figure}

\begin{figure}[htb] 
\begin{center}
\includegraphics[width=4cm]{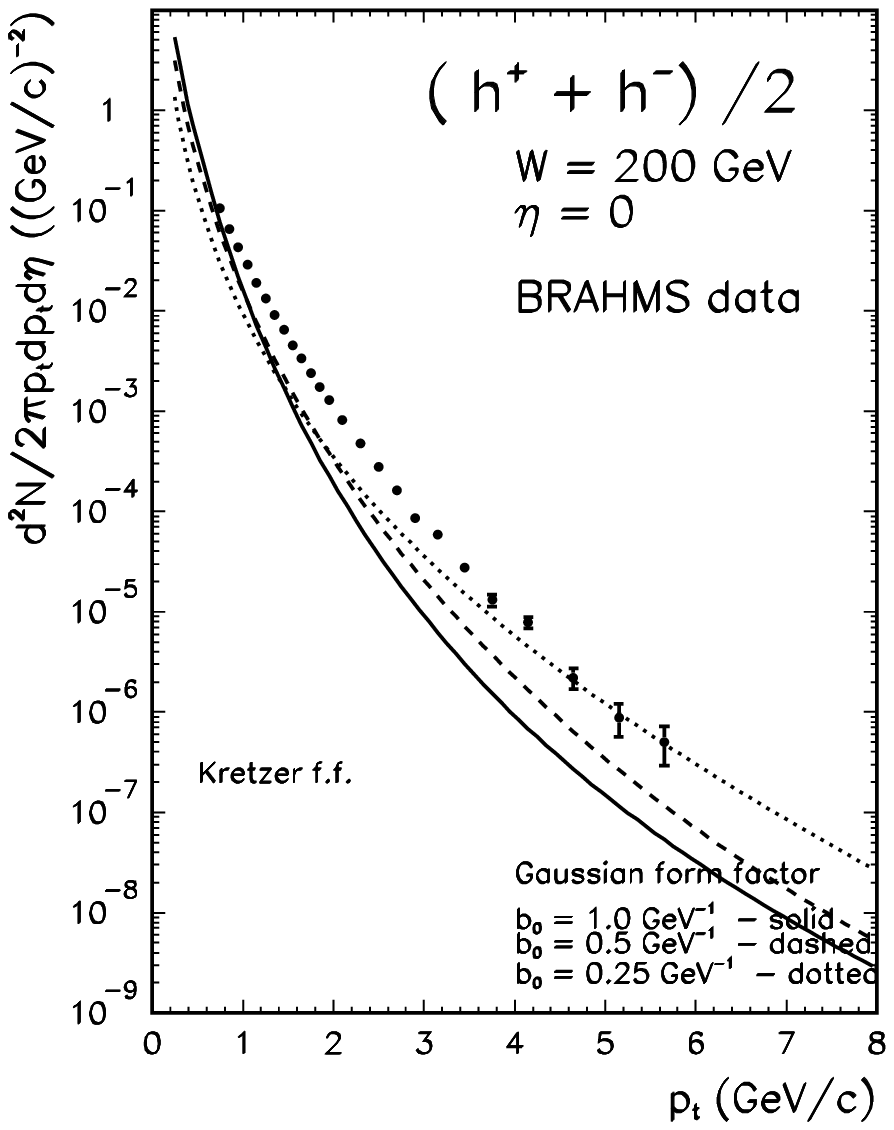}
\includegraphics[width=4cm]{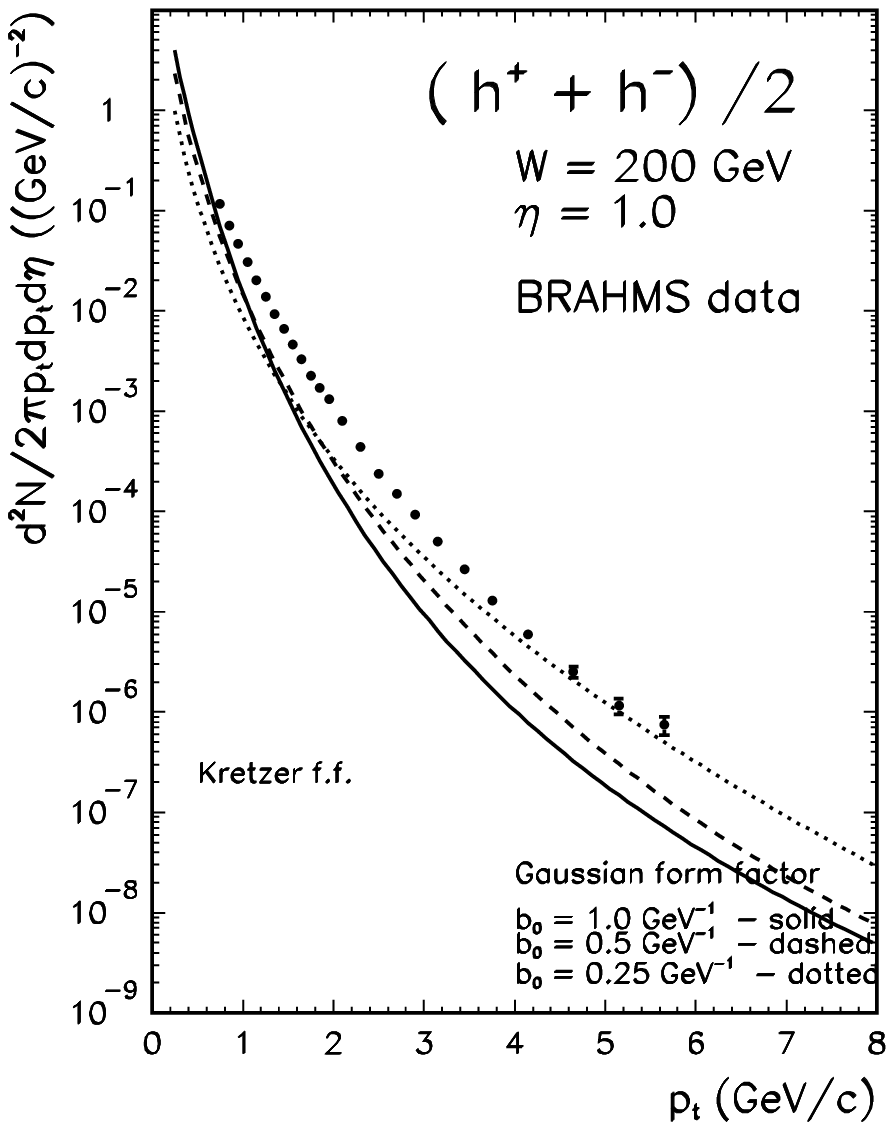}\\
\includegraphics[width=4cm]{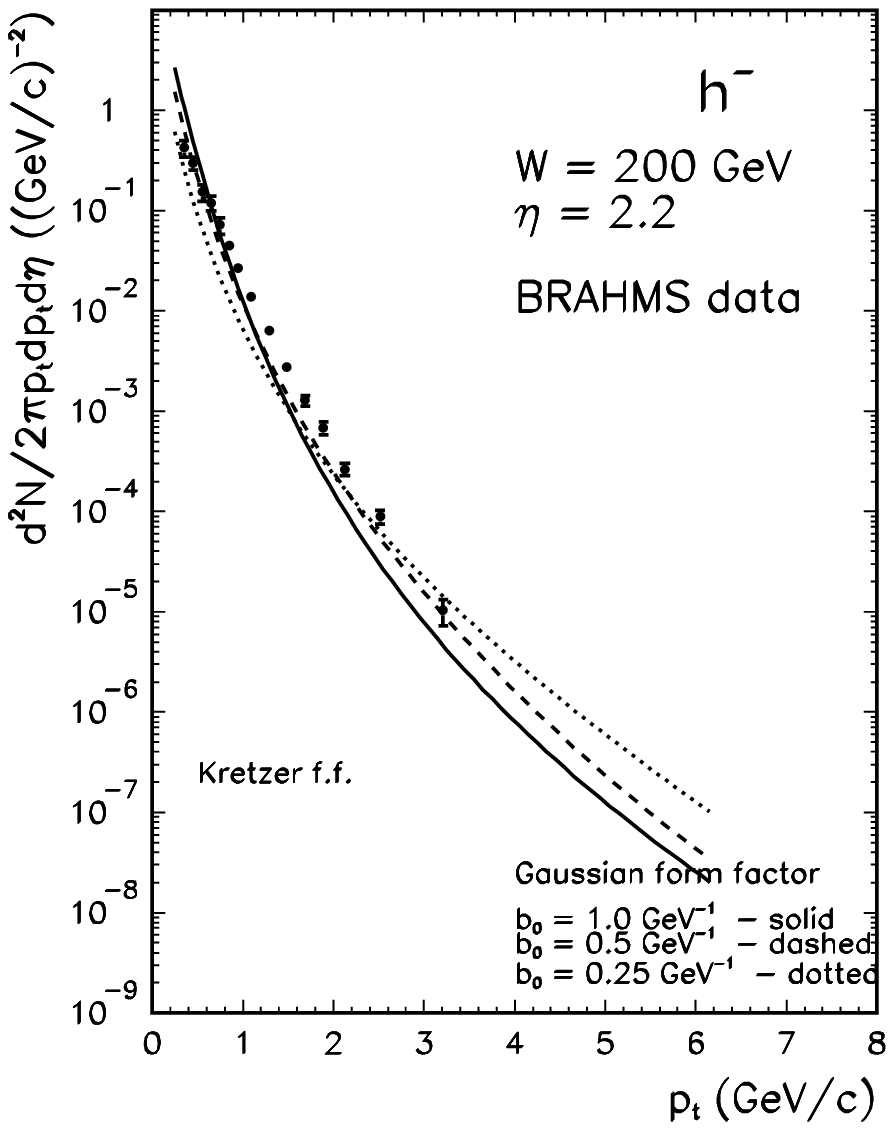}
\includegraphics[width=4cm]{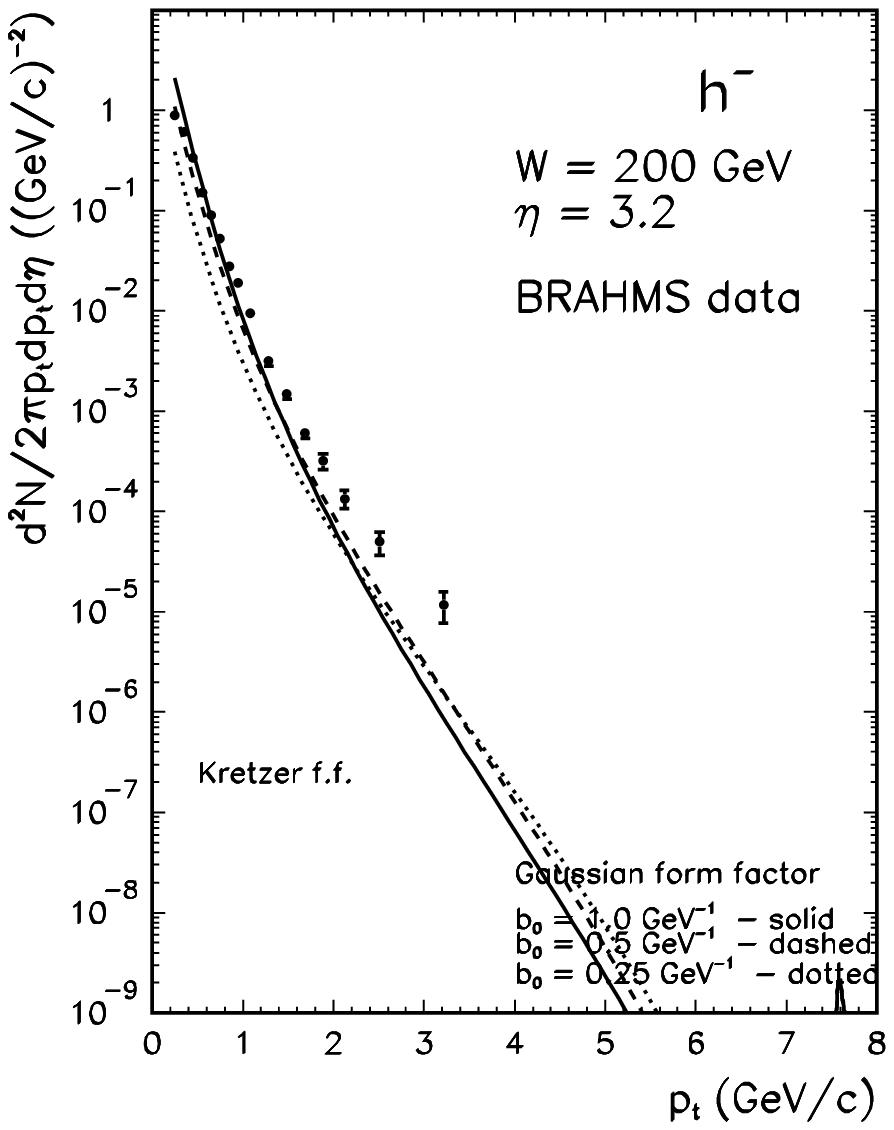}
\caption[*]{
Invariant cross section for charged particle production for
different values of particle pseudorapidity at W = 200 GeV.
The lines represent results obtained within our $k_t$-factorization
approach
for different values of parameter $b_0$ = 0.25, 0.5, 1.0 GeV$^{-1}$
and Kretzer fragmentation functions \cite{Kretzer00}.
The BRAHMS collaboration experimental data
\cite{BRAHMS_charged_hadrons} are shown
by the solid circles.
\label{fig:BRAHMS_our_b0}
}
\end{center}
\end{figure}

\begin{figure}[htb] 
\begin{center}
\includegraphics[width=4cm]{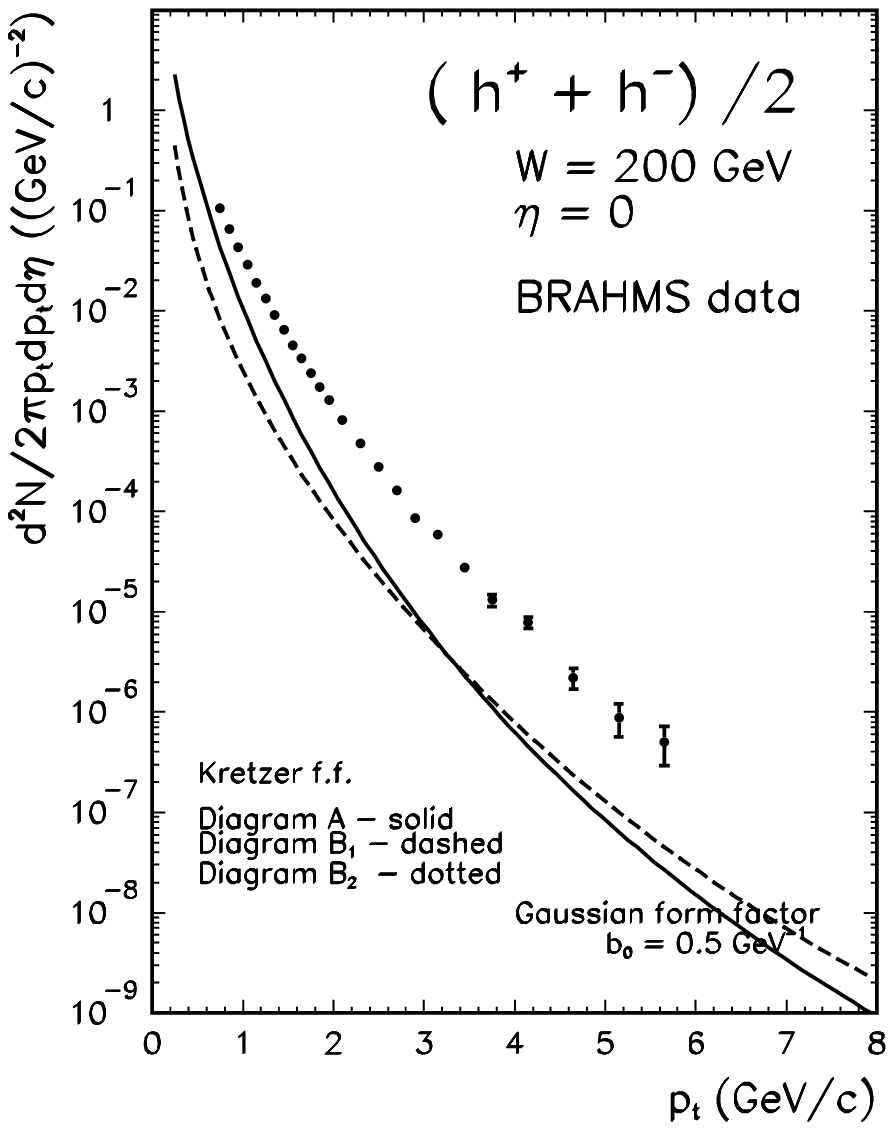}
\includegraphics[width=4cm]{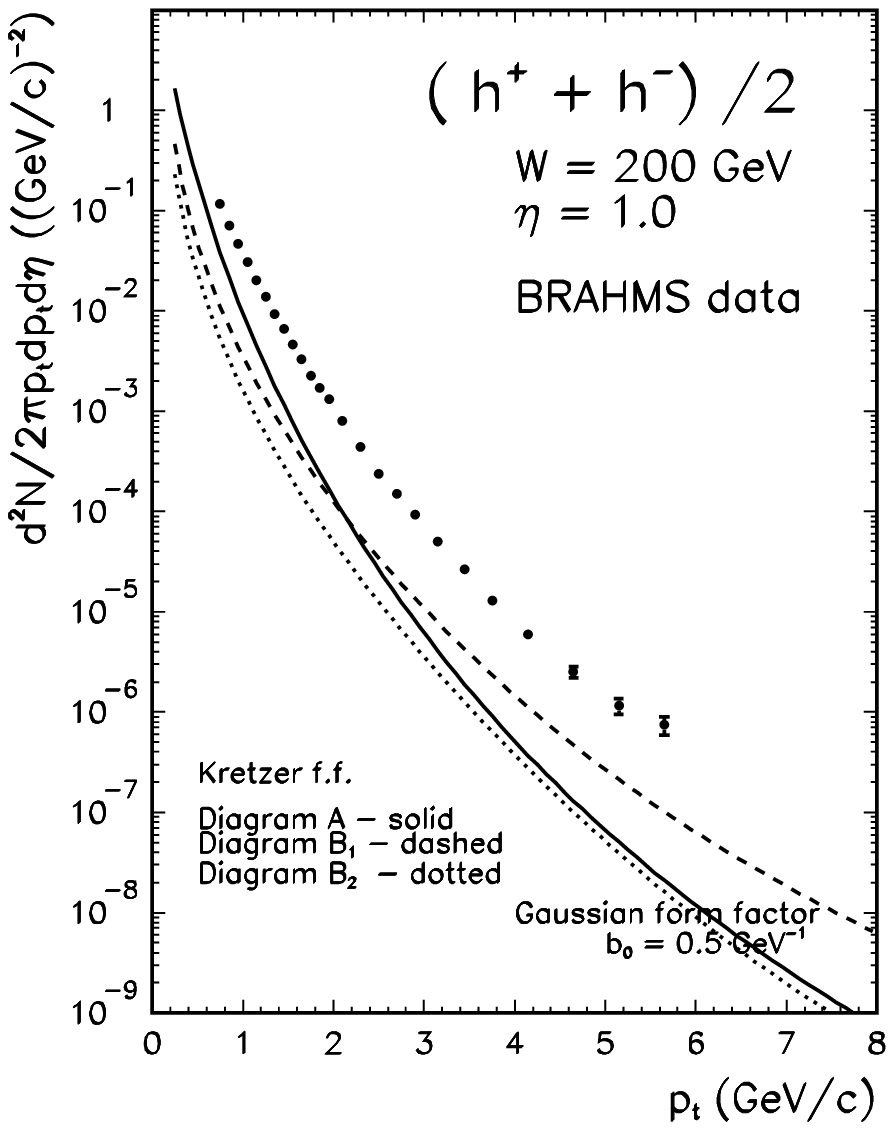}\\
\includegraphics[width=4cm]{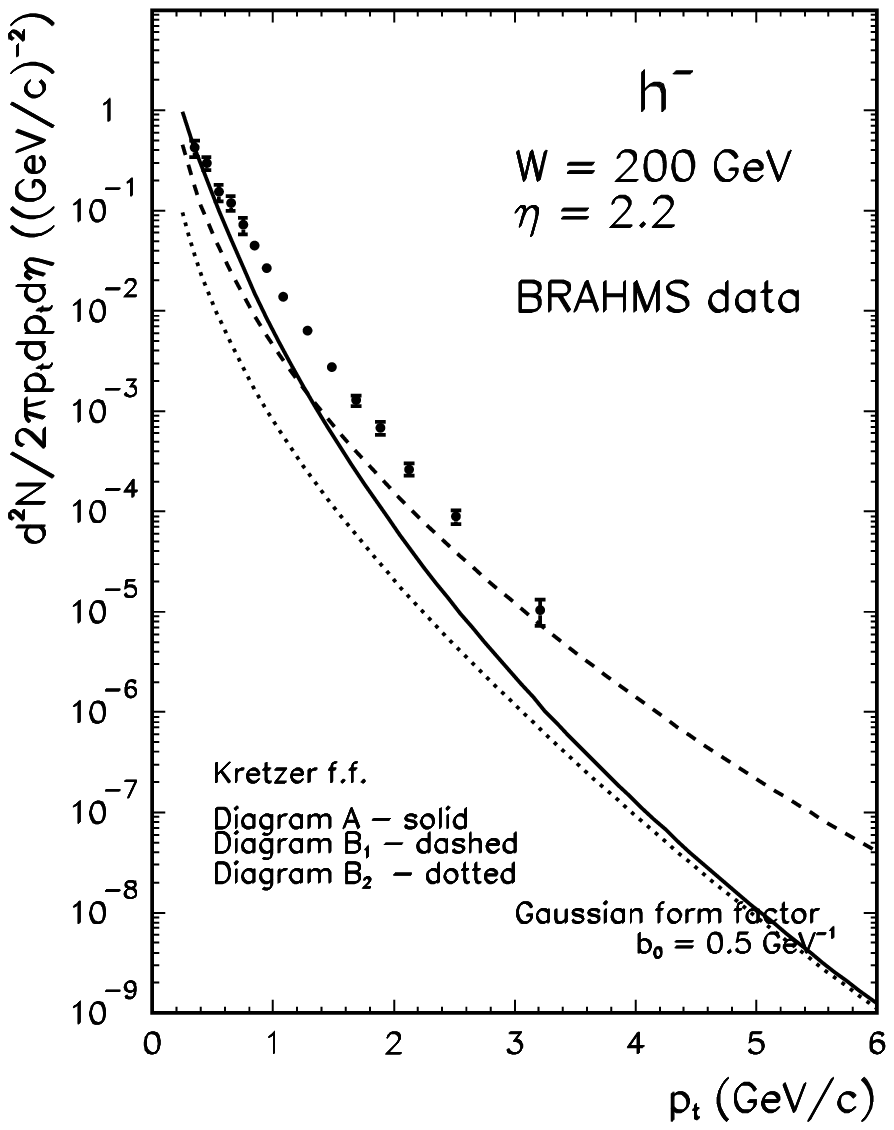}
\includegraphics[width=4cm]{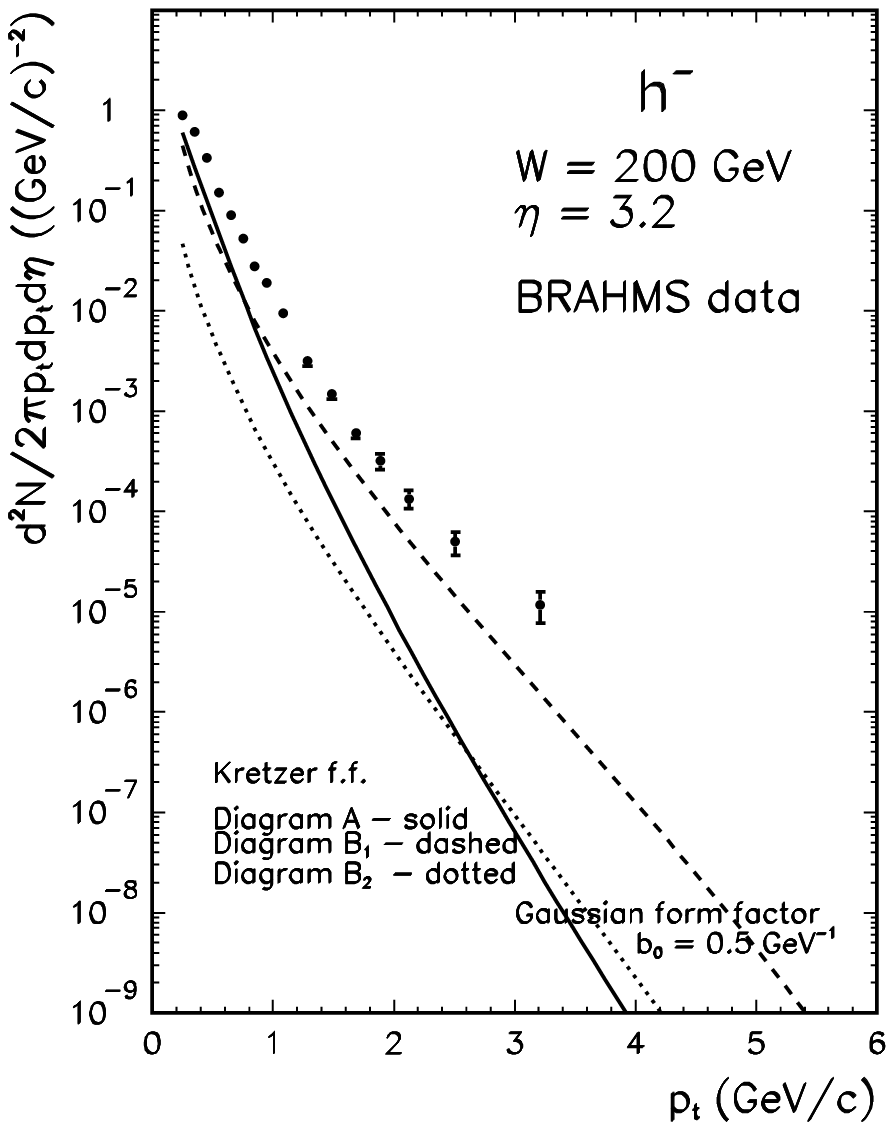}
\caption[*]{
Invariant cross section for charged particle production for
different values of particle pseudorapidity at W = 200 GeV.
The lines show individual contributions within our $k_t$-factorization
approach.
Here Kretzer fragmentation functions \cite{Kretzer00} were used.
The BRAHMS collaboration experimental data
\cite{BRAHMS_charged_hadrons} are shown
by the solid circles.
\label{fig:BRAHMS_our_contributions}
}
\end{center}
\end{figure}

Let us come now to charged particle distributions.
In Fig.\ref{fig:BRAHMS_coll} we show the results of the collinear approach
for different values of hadron pseudorapidity for the Kretzer
fragmentation functions \cite{Kretzer00}. The results of the
calculations are compared to the BRAHMS collaboration data
\cite{BRAHMS_charged_hadrons}. The theoretical calculations are
for the charged pions. The data are not exactly for pions but include all
charged hadrons, e.g. protons as well. 
There is a large factor of disagreement between the calculations
and the data. This factor seems to increase when going from
mid to forward rapidities which is not compatible with
the NLO collinear factorization approach where the so-called K-factor is
almost independent of rapidity.
In Fig.\ref{fig:BRAHMS_our_b0} we present result of our calculations
for $b_0$ = 0.5 GeV and the Kretzer fragmentation
functions \cite{Kretzer00}.
While at large (pseudo)rapidities ($\eta$ = 2.2, 3.2) our
results (negative pions) are quite compatible with the experimental
result for negative hadrons, there seems to be a missing contribution at
more central (pseudo)rapidities ($\eta$ = 0.0, 1.0), i.e. in the case when
the sum of positively and negatively charged hadrons is measured.
At present it is not clear to us if the missing strength is due
to protons and/or positively charged kaons. 
In Fig.\ref{fig:BRAHMS_our_contributions} individual contributions
from our $k_t$-factorization approach are shown.
They correspond to diagrams
$A$, $B_1$ and $B_2$ in Fig.1 of Ref.\cite{CS05}.
While at $\eta$ = 0 the $gg \to g$ contribution
dominates up to $p_t \sim$ 3.5 GeV at $\eta$ = 2.2 the $B_1$
contribution is larger than the $gg \to g$ one already at $p_t \sim$
1.5 GeV and at $\eta$ = 3.2 above $p_t \sim$ 1 GeV.
This is the $B_1$ contribution which provides a good description
of the BRAHMS data at forward ($\eta \gg$ 2) rapidities.
\begin{figure}[htb] 
\begin{center}
\includegraphics[width=6cm]{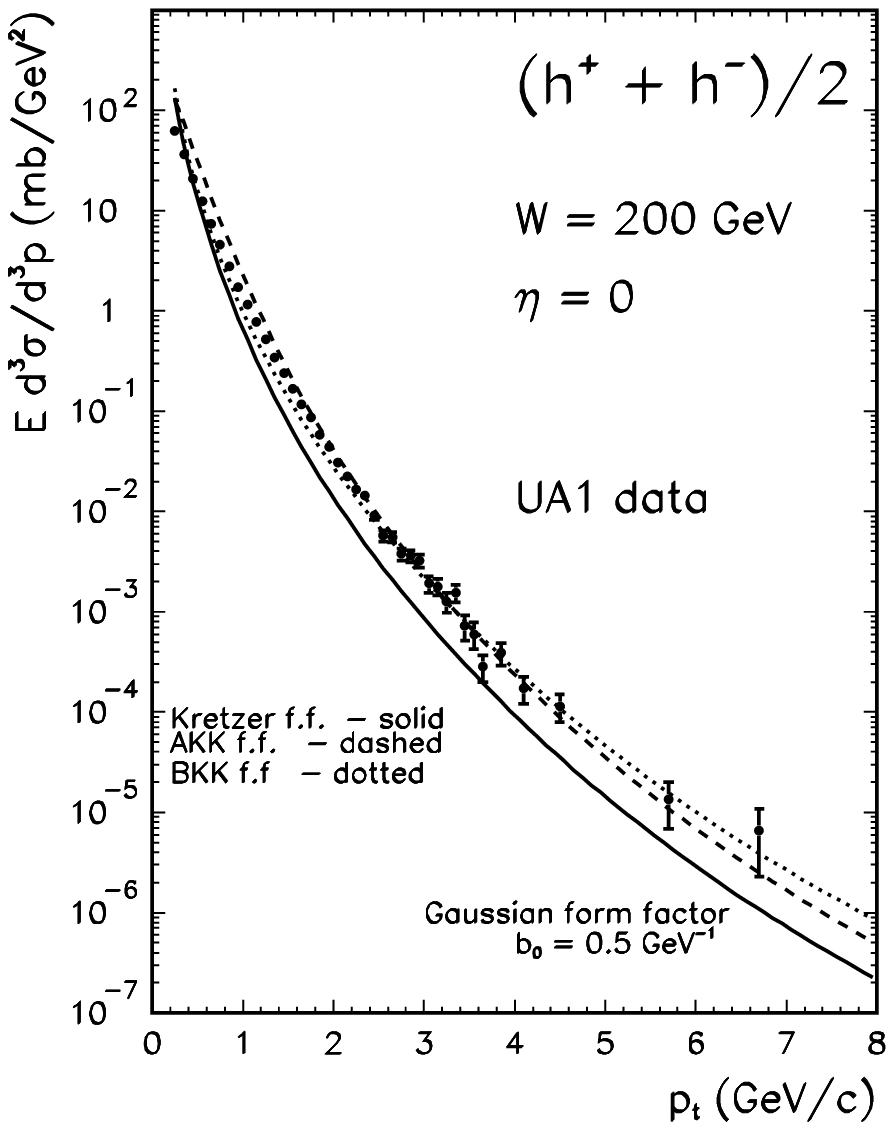}
\includegraphics[width=6cm]{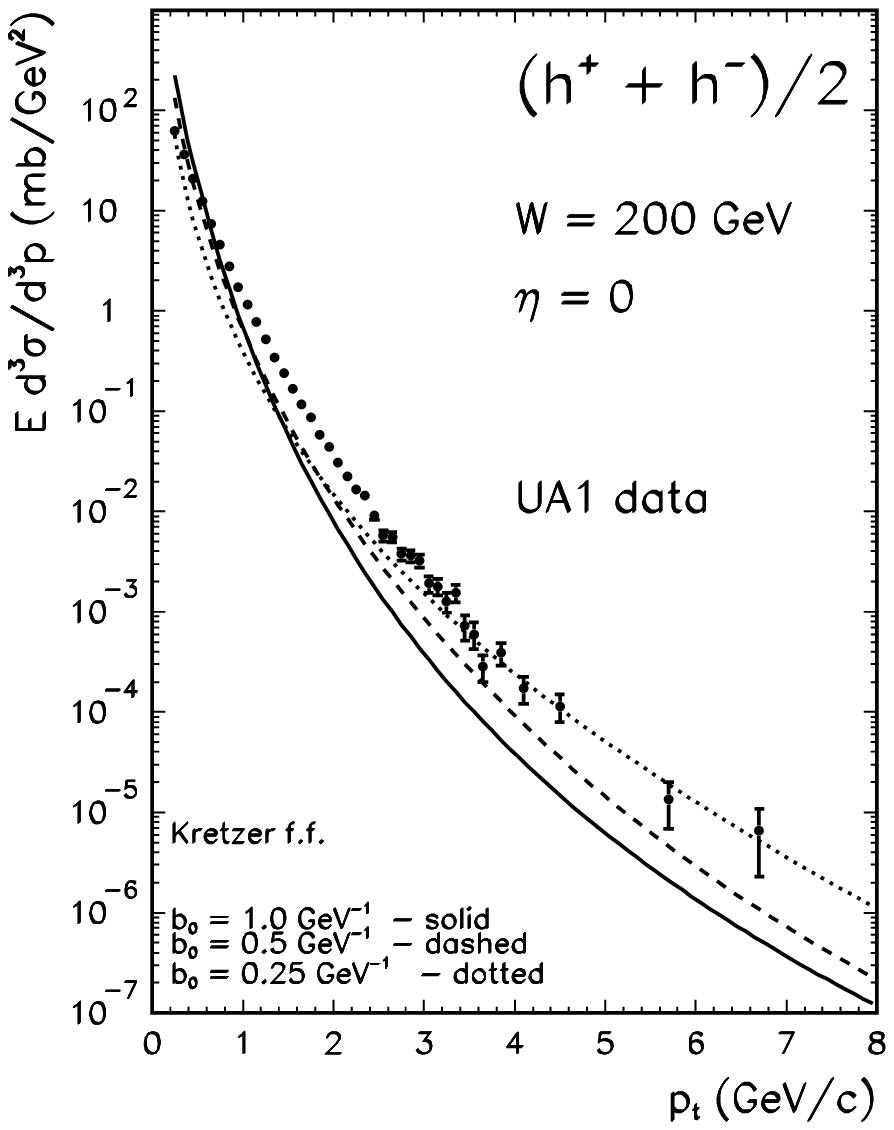}
\caption[*]{
Invariant cross section for charged particle production for
different values of particle pseudorapidity at W = 200 GeV.
The uncertainties on fragmentation function are discussed
in the left panel and the dependence on the $b_0$ parameter
in right panel.
The UA1 collaboration experimental data \cite{UA1_charged_hadrons}
are shown by the solid circles.
\label{fig:UA1_our}
}
\end{center}
\end{figure}
\begin{figure}[htb] 
\begin{center}
\includegraphics[width=6cm]{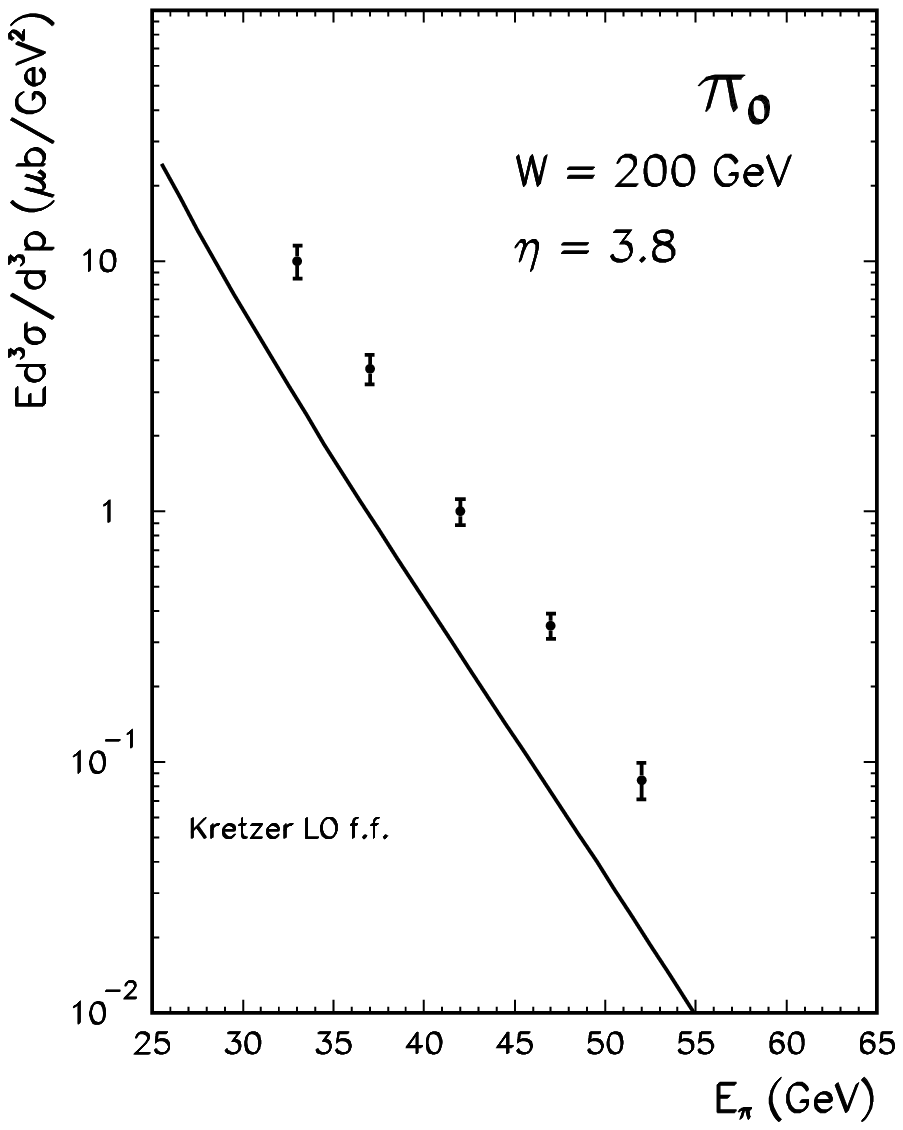}
\includegraphics[width=6cm]{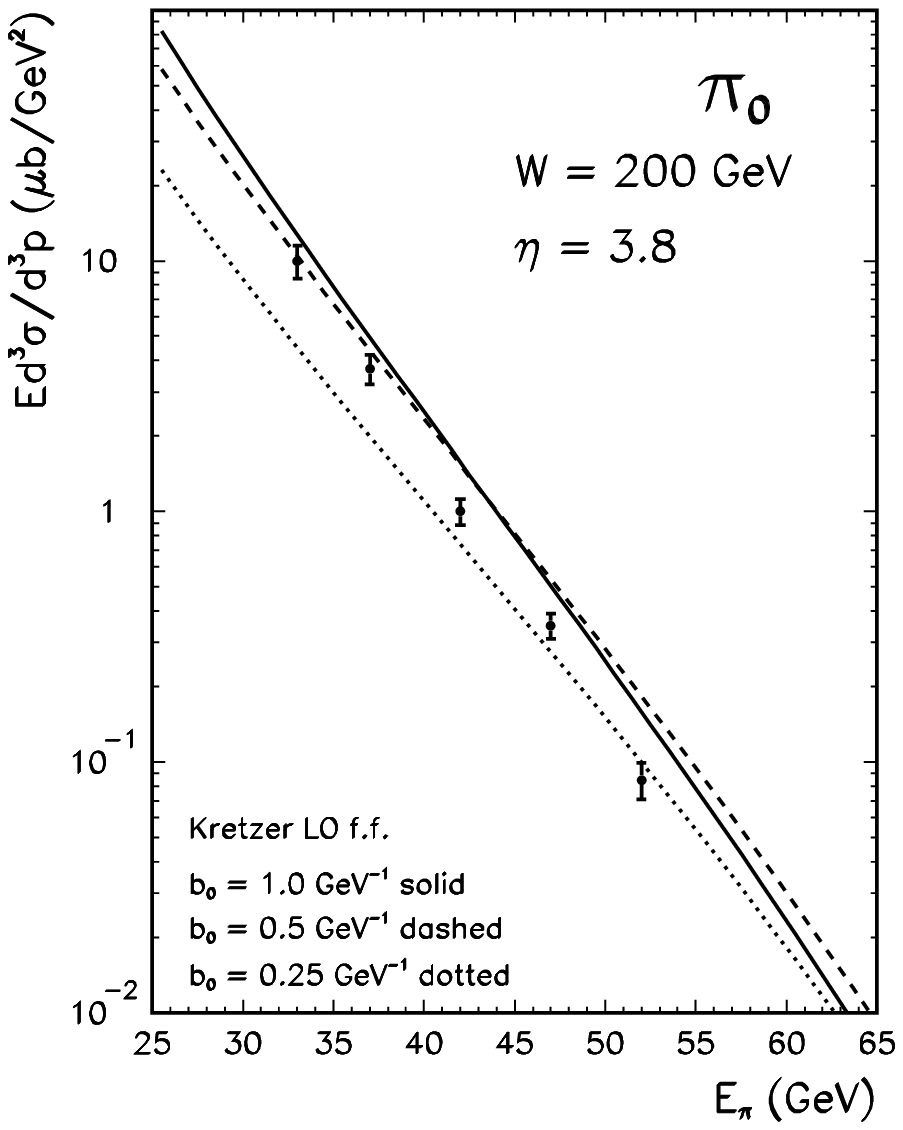}
\caption[*]{
Invariant cross section for neutral pion production as a function
of pion energy at W = 200 GeV for $\eta$ = 3.8 for collinear (left panel)
and our $k_t$-factorization (right panel) approach.
The STAR collaboration data are taken from Ref.\cite{STAR_pi0}.
\label{fig:STAR}
}
\end{center}
\end{figure}
\begin{figure}[htb] 
\begin{center}
\includegraphics[width=7cm]{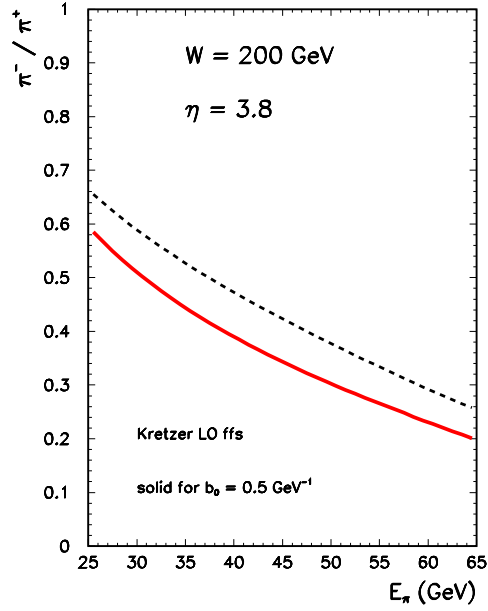}
\caption[*]{
The ratio of the $\pi^-$ to $\pi^+$ cross sections as a function
of pion energy for $\eta$ = 3.8 for collinear (dashed)
and our $k_t$-factorization (thick solid) approach.
\label{fig:pim_to_pip}
}
\end{center}
\end{figure}
\begin{figure}[htb] 
\begin{center}
\includegraphics[width=7cm]{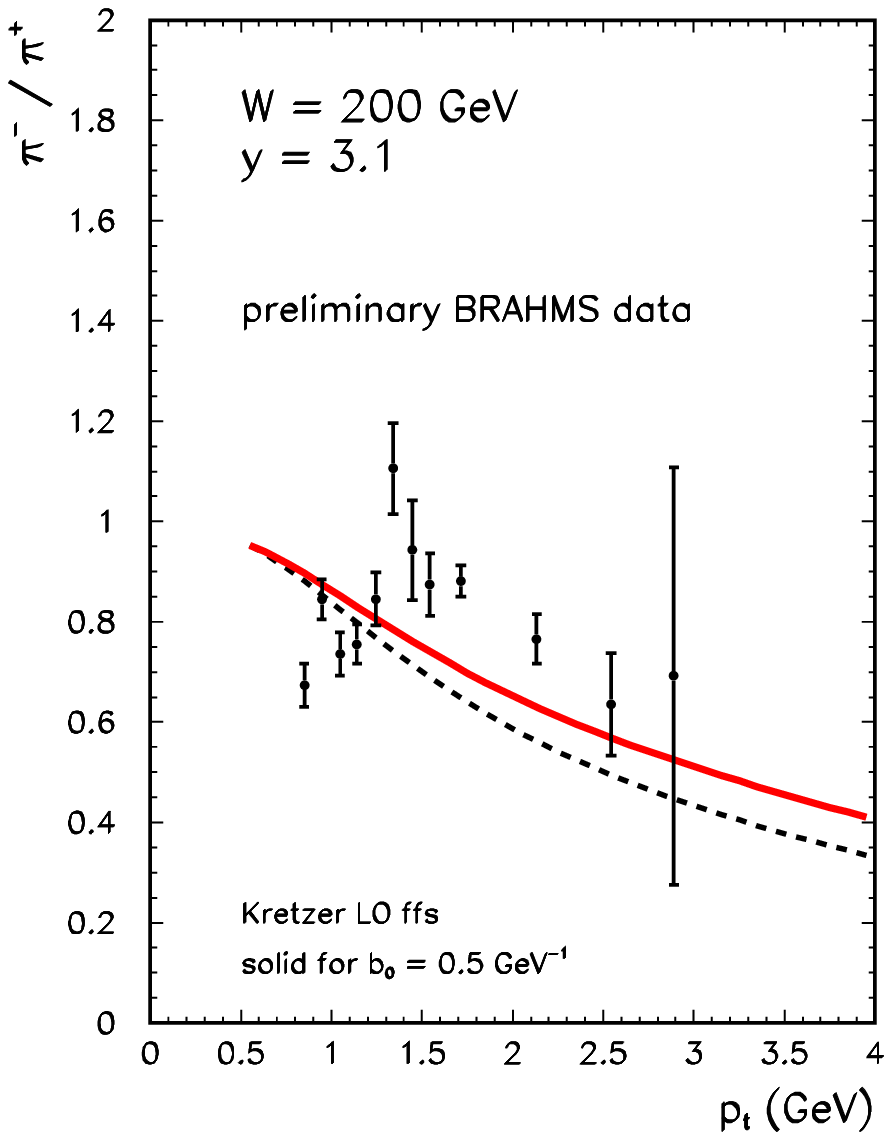}
\caption[*]{
The ratio of the $\pi^-$ to $\pi^+$ cross sections as a function
of pion transverse momentum for y = 3.1 for collinear (dashed)
and our $k_t$-factorization (thick solid) approach.
The data are from a recent presentation at Quark Matter 2005 conference
\cite{Karabowicz_talk}.
\label{fig:pim_to_pip_y3_1}
}
\end{center}
\end{figure}

In Fig.\ref{fig:UA1_our} we compare results of our calculations
with the old proton-antiproton charged hadron data
\cite{UA1_charged_hadrons} for $\eta$ = 0.
The dependence on fragmentation functions is shown
in the left panel. In the right panel we show the dependence
on the parameter $b_0$ for the Kretzer fragmentation functions.
The figure looks very similar to the figure with the BRAHMS
collaboration data for $\eta$ = 0, which simply reflects consistency
of the proton-proton (BRAHMS) data and proton-antiproton (UA1) data.
Within the approximations used in our approach the charged
pion inclusive cross section is identical for the proton-proton
and proton-antiproton collisions, provided the energy is the same,
which is the case for the BRAHMS and UA1 colaboration data.

Let us concentrate now at the very forward region of the phase space.
Recently the STAR collaboration has published \cite{STAR_pi0}
large-rapidity, intermediate-transverse-momentum
data for $\pi^0$ production.
In Fig.\ref{fig:STAR} we compare the results of the collinear (left panel)
and our $k_t$-factorization approach (right panel) calculated
with the Kretzer fragmentation functions.
While the collinear approach underestimates the STAR experimental
data by a factor of about 3, the $k_t$-factorization approach
is almost consistent with the data, especially at smaller pion
energies, i.e. at not too high transverse momenta. The situation would
change somewhat with different set of fragmentation functions.

In our approach the diagrams with quark degrees of freedom
($B_1$ and $B_2$) lead to an
asymmetry in $\pi^+$ and $\pi^-$ production.
Recent results of the BRAHMS collaboration for the transverse momentum
integrated cross section $d\sigma/dy$ show the $\pi^+ - \pi^-$
asymmetry already above y=2 \cite{BRAHMS_pp_asymmetry}.
As an example in Fig.\ref{fig:pim_to_pip} we show the ratio of
the corresponding cross sections for $\pi^-$ and $\pi^+$ production for
the STAR kinematics. A huge deviation from the unity can be observed.
The $k_t$-factorization approach ratio (thick solid line)
is somewhat smaller than the corresponding ratio in the collinear
approach (dashed line).

During the preparation of this manuscript the BRAHMS collaboration
has presented the first preliminary data for identified $\pi^+$ and
$\pi^-$ \cite{Karabowicz_talk}.
In Fig.\ref{fig:pim_to_pip_y3_1} we show
transverse momentum dependence
of the ratio $\pi^- / \pi^+$ for rapidity y = 3.1.
Only statistical error bars are shown. Although the experimental
ratio is not completely monotonous it clearly shows a deviation
from unity which is an experimental evidence that gluon degrees of
freedom are not sufficient to describe the production of hadrons.
We describe the main trend of deviation of
the ratio from unity relatively well except in the region of very
small values of transverse momenta.
\footnote{We expect that this region is sensitive to
pion stripping \cite{pion_stripping}
as well as diffractive production of mesons not included here
explicitly.}
Of course such a ratio depends on the details of the fragmentation
functions and in particular on their flavour decomposition
which as discussed above is very difficult to obtain from
the $e^+ e^-$ scattering alone.
We hope that in the near future the BRAHMS collaboration will be able
to scan the $\pi^- / \pi^+$ ratio as a function of (pseudo)rapidity and
transverse momentum in order to identify the contributions with
quark degrees of freedom and perhaps to extract flavour-dependent
fragmentation functions.

In nuclear collisions the $\pi^+ - \pi^-$ asymmetry is weakened
by the presence of the $p p$, $n n$, $p n$ and $n p$ subcollisions.
Due to isospin symmetry relation, this leads to equal yield
of $\pi^+$ and $\pi^-$ for collisions of isospin symmetric
nuclei. For collisions of heavy nuclei there is a small nonzero effect
due to the excess of neutrons over protons.
The asymmetry can be, however, quite sizeable in peripheral
collisions \cite{PS04} due to neutron skin effects.
The sign of the nuclear asymmetry is then reversed as compared to
the proton-proton collisions.
The small asymmetry in nuclear collisions was recently mistakenly
interpreted as a dominance of purely gluonic effects even
at large rapidities. Our calculation actually shows that
the quark terms dominate at very forward/backward rapidity regions.
Although our calculation is for elementary proton-proton collisions
only it puts into question some recent nuclear color glass condensate
calculations based on gluon degrees of freedom only.

\section{Conclusions}

We have shown that the formalism recently developed by us and based on
unintegrated parton distributions which fulfill the so-called
Kwieci\'nski evolution equations provides a reasonable description
of the recent experimental data of the PHENIX, BRAHMS and STAR
collaborations at RHIC. The description is particularly good
in the region of intermediate transverse momenta of pions
$p_{t,h} \sim$ 1 -- 4 GeV.
In comparison, the standard collinear factorization approach gives
results which are by a factor of 3 -- 7 lower than the experimental data,
depending on particle rapidity.
We have found a rather strong dependence on the set of fragmentation
functions used in the calculation.

Inclusion of diagrams with quark degrees of freedom leads
to $\pi^+ - \pi^-$ asymmetry. The preliminary BRAHMS data
provide evidence for such an asymmetry.
A dedicated measurement of the $\pi^+ - \pi^-$ asymmetry in
forward and backward region as a function of transverse momentum
would be a good test of the present approach
and perhaps could be used to constrain better the gluon-to-pion
and quark-to-pion fragmentation functions which extraction in
$e^+ e^-$ collisions is ambigous to a large extent.

\vskip 1cm

{\bf Acknowledgements}
We are indebted to Rados{\l}aw Karabowicz for providing us
with preliminary results of the BRAHMS collaboration for $\pi^+$ and
$\pi^-$ production presented at Quark Matter 2005.
We are also indebted to Peter Levai for an interesting discussion
about collinear approach and Stefan Kretzer for a discussion about
parton fragmentation functions.
This work was partially supported by the Polish KBN
grant no. 1 P03B 028 28.



\end{document}